 \documentclass[preprint2]{aastex}

\shorttitle{A catalogue of low mass galaxies with central candidate black holes}
\shortauthors{Nucita et al.}

\usepackage[utf8]{inputenc}
\usepackage{graphicx,epsfig}  %! ,mathabx}    % marvosym}
\usepackage{amsmath,amssymb}
\usepackage{subfigure}
\usepackage{longtable}
\usepackage{lscape}
\usepackage{times}
\usepackage{textcomp}
\usepackage{natbib}
\usepackage{txfonts}
\usepackage{multirow}
\usepackage{fixltx2e}

\def\ut#1{\mathop{\vtop{\ialign{##\crcr
     $\hfil\displaystyle{#1}\hfil$\crcr\noalign
     {\kern1pt\nointerlineskip}\hbox{$\hfil\sim\hfil$}\crcr
     \noalign{\kern1pt}}}}}

\def\undersymbol#1#2{\mathop{\vtop{\ialign{##\crcr
     $\hfil\displaystyle{#2}\hfil$\crcr\noalign
     {\kern1pt\nointerlineskip}\hbox{$\hfil#1\hfil$}\crcr
     \noalign{\kern1pt}}}}}
\def\arcsec{^{\prime\prime}}

\def\degr{^\circ}

\begin{document}

% low mass galaxy in x ray 

\title{A catalogue sample of {low mass} galaxies observed in $X$-rays with central candidate black holes}
 \author{A. A. Nucita, L. Manni, F. De Paolis, M. Giordano, and G. Ingrosso}
 \affil{Department of Mathematics and Physics {\it ``E. De Giorgi''}, University of Salento, Via per Arnesano, CP 193, I-73100,
  Lecce, Italy}
 \affil{INFN, Sez. di Lecce, Via per Arnesano, CP 193, I-73100, Lecce, Italy\\
Corresponding author:   \email{nucita@le.infn.it}}

\begin{abstract}
We present a sample of $X$-ray selected candidate black holes in 51 {low mass} galaxies with $z\le 0.055$ {and mass up to $10^{10}$ M$_{\odot}$} 
obtained by cross-correlating the NASA-SLOAN Atlas with the 3XMM catalogue. 
{We have also searched in the available catalogues for radio counterparts of the black hole candidates 
and find that 19 of the previously selected sources have also a radio counterpart.} 
Our results show that about $37\%$ of the galaxies of our sample 
host {  an $X$-ray source} (associated to a radio counterpart) spatially 
coincident with the galaxy center,  in agreement with {  other recent works}. 
For these {\it nuclear} sources, the $X$-ray/radio fundamental plane relation allows one to 
estimate the mass of the (central) candidate black holes which results to be in the range
$10^{4}-2\times10^{8}$ M$_{\odot}$ ({with median value of $\simeq 3\times 10^7$ M$_{\odot}$ and eight candidates 
having mass below $10^{7}$ M$_{\odot}$}). This result, while suggesting  that 
$X$-ray emitting black holes in low-mass galaxies may have had a key role in the evolution of such systems, makes even more urgent 
to explain how such massive objects formed in galaxies. {Of course, dedicated follow-up observations both in the $X$-ray and radio bands, as well as in the optical, 
are necessary in order to confirm our results.}

\end{abstract}


\section{Introduction}
{Low mass galaxies with stellar mass less than $\simeq 1\times 10^{10}$ M$_{\odot}$ are faint galaxies, 
particularly hard to be detected and studied. The objects with mass below $\simeq 5\times 10^{9}$ M$_{\odot}$ are usually classified as isolated or satellite dwarf galaxies  
being this value the visible mass of the Large Magellanic Cloud \citep{gyuk,conroy}.

Low mass galaxies span all the possible shape classification from spirals to ellipticals through irregular objects\footnote{ 
In this context, dwarf spheroidal galaxies are a sub-class of low mass galaxies with {stellar mass in the range $10^3$ - $10^{7}$} M$_{\odot}$ 
(\citealt{martin2008}) particularly interesting since they show large mass-to-light ratios which {make them to
be dominated} by dark matter (\citealt{mateo1997}). The interest in dwarf galaxies is rapidly 
growing (see, e.g., \citealt{McConnachie2012}) both for stellar populations studies (\citealt{amorisco2012,maccarone2005b}) and 
searches for central IMBHs (see e.g., \citealt{reines2013}, \citealt{nucita2013a}, \citealt{nucita2013b}, \citealt{manni2015}).}.}

It is known that massive black holes\footnote{The paradigm of the existence of massive black holes in the center of almost all galaxies comes from
indirect observations at several wavelenghts. In the {  next future}, the Event Horizon Telescope (which is based on techniques of Very Long Baseline Interferometry -VLBI- see, e.g., \citealt{eht}) will achieve the $\mu$-arcseconds angular resolution
necessary to resolve the shadows of massive black holes at least in nearby galaxies (see, e.g., \citealt{melia} for the Sgr A* case). In these cases, the shape of such shadows will allow one not only to have the first direct evidence of the existence of such objects but also to get 
an estimate of their mass, spin and charge parameters 
(see, e.g., \citealt{nucita2007,nucita2011,zak2012}).} are hosted in the nuclei of almost every galaxy characterized by a central bulge (\citealt{kormendyho2013}). In these cases, black holes reveal 
themselves via stellar and gas kinematic in close targets or, in distant active galactic nuclei, by mean of the emitted radiation. 
By extrapolating the fundamental  $M_{BH} - M_{Bulge}$ relation  (\citealt{magorrian1998}, for the super massive BH case but see also \citealt{reines2015}) down to the typical mass of low mass galaxies, one
realizes that intermediate mass black holes (IMBHs) may also be found in such stellar systems. Their number density and characteristics is, in turn, of crucial importance to get information about the black hole 
seed population in the early Universe (\citealt{volonteri2010,natarajan2014}) and about their contribution in re-ionizing the hydrogen at high red-shift via the $X$-ray emission (\citealt{volonterignedin2009}).  Furthermore, 
searching for the high energy emission expected from these objects and applying the fundamental plane relation  at radio and $X$-ray wavelengths (see \citealt{merloni2003}) as recently done by several authors 
(see e.g., \citealt{reines2013}, \citealt{nucita2013a}, \citealt{nucita2013b}) allows one to infer the mass of the compact object (if any). 

Recently, \citet{lemons2015} cross-matched a sample of 44000 {dwarfs} (provided by the NASA-Sloan Atlas\footnote{\tt http://www.nsatlas.org.})  
with mass within $3\times 10^9$ M$_{\odot}$ and at a redshift as large as $z<0.055$  with the Chandra Source Catalogue (\citealt{evans2010}) finding a heterogeneous sample of 19 galaxies with a total of
43 point-like $X$-ray sources with hard spectra and $2-10$ keV luminosities between $10^{37}$ erg s$^{-1}$ and $10^{40}$ erg s$^{-1}$, i.e. in the typical 
range of power emitted by stellar-mass $X$-ray binaries and massive black holes accreting at low Eddington rate. Of these sources, the author set an upper limit of 53\% 
on the fraction of galaxies in the sample having a hard $X$-ray source located in the optical nucleus, although this region is generally poorly constrained for 
most of the dwarfs. {Furthermore, \citet{nucita2013a,nucita2013b,manni2015} have focused their attention on a sample of five dwarf spheroidal Milky Way satellites (i.e., Fornax,  
Ursa Minor, Draco, Leo I, and Ursa Major II) since, due their proximity and the lack of overcrowded environment, allows one to characterize (on statistical basis) the nature of the identified $X$-ray sources. In particular, it were 
found hints of the existence of central massive black holes in $\simeq 40\%$  of the cases, but it was also derived a small contamination due to a few local stellar-mass $X$-ray binaries 
so that follow-up observations are required to disentangle between the stellar-mass $X$-ray binary and  active galactic nucleus with an accreting black hole scenarios.}

In this paper, we search for the signatures of massive black holes in low mass galaxies by using archival 
data from the {\it XMM}-Newton satellite. In spite of its low angular resolution ($\simeq 6\arcsec$, primarily 
due to the point spread function - PSF - of the mirror modules, see e.g. \citealt{xmmhandbook,xrps}),  {\it XMM}-Newton 
is particularly useful for studying faint objects (as the low accreting/emitting massive black holes targets of this study) 
thanks to its large effective area (\citealt{jansen2001}). We cross-correlated the NASA-Sloan Atlas with the 3XMM-DR5 catalogue (\citealt{rosen2016}), which is five times the current size of the Chandra source catalogue 
(\citealt{evans2010}) producing a starting galaxy sample. For each target, we retrieved and analyzed the original {\it XMM}-Newton  data products (ODFs),  produced images in the {\it $0.2-12$} keV energy band and 
accumulated spectra for each of the matched sources. 

{Our catalogue (hereafter GiX, i.e. galaxies in $X$-rays) resulted in 51 galaxies {detected by using 
the {\it XMM}-Newton satellite. The corresponding masses turn out to span between about 
$1.5\times 10^8$ and $10^{10}$ M$_{\odot}$, with $5.4\times 10^9 M_\odot$ as the median value.}
A further cross correlation of  
GiX with FIRST\footnote{The FIRST catalogue is available at {\tt http://sundog.stsci.edu/}.} 
(the VLA FIRST Survey: Faint Images of the Radio Sky at Twenty cm), {NVSS\footnote{The NVSS catalogue is 
available at {\tt  http://www.cv.nrao.edu/nvss/}.} (the NRAO VLA Sky Survey), and the 
catalogue of $X$-ray selected star-forming galaxies within the {\it Chandra Deep Field}-South by \citet{rosagonzalez2007} (hereafter, RG2007)} resulted in 19 
radio counterparts allowing us to use the fundamental plane relation  (see \citealt{merloni2003}) and infer the black hole mass.}

This paper is organized as follows. In Sect. \ref{sample} our main galaxy sample is outlined. In Sect. \ref{xraydatanalysis} we give details on the data analysis of the {\it XMM}-Newton data. 
In Sect. \ref{correlation} we present the cross-correlation among the GiX and FIRST, NVSS and RG2007 catalogues and use the fundamental plane relation to estimate the black hole masses. Finally, 
in Sect. \ref{results}, we address our results.

\section{The GiX sample}
\label{sample}
The NASA-Sloan Atlas is a catalogue of local galaxies ({145155 objects} up to redshift $z\simeq 0.055$) based on  reanalysis 
of optical and ultraviolet observations conducted for the SDSS and Galaxy Evolution Explorer (GALEX). Querying the catalogue
results in various galaxy parameters as the distance and mass estimate based on the {\it kcorrect} code of \citet{blanton2007}.  In particular, the mass
of each galaxy is  given in units of $M_{\odot} h^{-2}$ and we adopt $h=0.73$ {(see also the next Section)}. 

{We decided to consider galaxies with masses $\ut < 10^{10}$ M$_{\odot}$, 
i.e. {  a factor 3.3 larger than what assumed in \citet{lemons2015} (but see also \citealt{mezcua2016,pardo2016}). Hence, we selected 82977 galaxies from the 
NASA-Sloan Atlas, a number which reduces to 44594 objects with mass up to $3\times 10^{9}$ M$_{\odot}$, consistent with \citet{lemons2015}. Note that under these assumptions, our 
selected objects would not be properly classified as dwarf galaxies.} 
 
{We do not set a lower limit for the stellar mass because the parental Sloan catalogue contains objects with mass greater than $1\times10^7$ M$_{\odot}$
due to the $r<17$ spectroscopic apparent magnitude limit of SDSS.} As in \citet{reines2013}, we further put some constraints on the spectroscopic values 
of the SDSS sample. In particular, we selected only sources with EW $>$ 1 {A} and S/N $\geq$ 3 for $H\alpha$, $[N\,II]$ and $[O\,III]$,
whereas S/N $\geq$ 2 for $H\beta$.  {Finally, our parent low mass galaxy sample is constituted by 47959 sources}\footnote{  
When we restrict the upper value of the galaxy mass to $3\times 10^9$ M$_{\odot}$ we find 25974 objects which is exactly 
the number found by \citet{reines2013} when adopting the same selection criteria.}.

We then searched for galaxy counterparts in $X$-rays in the {recent fifth}
release of the 3XMM catalogue (3XMM-DR5, \citealt{rosen2016}), the 
largest $X$-ray source catalogue ever produced. For each detected sources, one has the fluxes and count rates in 7 $X$-ray energy bands, the total $0.2-12$ keV band counts, 
and four hardness ratios, as well as the associated celestial coordinates and a measure of the detection quality (controlled via the 
the detection summary flag {\it SUM$\_$FLAG}). We only considered sources characterized by a detection quality of $0$ (good sources) or $1$ (if at least one of the warning flags 
- indicating low detector coverage, proximity to other sources, within extended emission, and/or near bright corner of CCD - was activated, but no possible-spurious-detection flags were on). 
We verified that the interesting sources of our sample have a maximum detection likelihood value (defined as $-ln(P)$, where $P$ is the probability of the detection occurring by chance) {  larger than $\simeq 6.7$.}
In the association procedure, a galaxy of our sample was considered cross-matched with a corresponding source in the 3XMM-DR5 catalogue if their distance was less or equal than 
$3\arcsec$. 

For each of the interesting source, we retrieved the {\it XMM}-Newton raw data files (ODF) and produced $X$-ray images in the {\it $0.2-12$} keV band (see next Section for details on the data reduction pipeline). 
We visually inspected the images in order to eliminate faked sources. Hence, our final
GiX catalogue is constituted by 51 targets with 40 sources characterized by $0$ detection quality flag, and 11 with this flag set to 1. In Table \ref{tableobsidxmm}, we give a sequential number, 
the {\it XMM}-Newton OBSID, the NASA-Sloan Atlas ID, the common galaxy name, the target coordinates (J2000 RA and DEC in degrees), the distance (in arcseconds) between the galaxy center and the cross-matched $X$-ray source, 
the galaxy mass according to the NASA-Sloan Atlas, the SUM$\_$FLAG {  (labeled as SF)} value characterizing the quality of the $X$-ray observation, {and the relative {net counts} along with the corresponding error,} respectively. 
{It is worth noting that all of our sources (except the source $\# 36$) seem to be point-like as suggested by the extent parameter derived by the SAS task {\it emldetect} and used 
when compiling the 3XMM catalogue. In the case of the aforementioned source $\# 36$, the associate extension is $\simeq 7.4\arcsec$ prompting to a non point-like origin and possibly related to a non resolved 
source group. In this context, the black hole mass for source $\# 36$ estimated via the $X$-ray/radio fundamental plane (see next) has to be considered as a lower limit.}

\section{$X$-ray data analysis}
\label{xraydatanalysis}
{  In this paper, we used an ab-initio procedure in order to get estimates of the source fluxes in the $0.2-12$ keV and $2-10$ keV bands and not relied blindly 
on the data available on the 3XMM-DR5 catalogue. The observation data files (ODFs) were processed using the XMM-Science Analysis System (SAS version 14.0.0) together with the most updated calibration constituent files. }

The event lists for the three EPIC cameras were obtained by using the standard {\it emchain} and {\it epchain} tools following standard procedures for the screening part (\citealt{xrps}).
{  In particular, we rejected time intervals characterized by high levels of background activity. In this respect, we constructed light curves in the energy range $10-15$ keV with a 
given bin size, evaluated  the mean and the standard deviation $\sigma$ of the time series and cut the time intervals with a number of counts per bin larger than $3\sigma$.  We iterated the whole procedure until 
the number of counts per bin is constant\footnote{  A similar approach giving comparable results is described in {\tt http://www.sr.bham.ac.uk/xmm2/}.}. In presence of observations affected by large flares, we used a more 
restrictive cut by simply requiring to have a threshold of 0.4 counts s$^{-1}$
and 0.35 counts s$^{-1}$ for the pn and MOS cameras, respectively.}

We produced images in the {\it $0.2-12$} keV band for the three cameras visually searching for point-like sources at the nominal coordinates of the galaxies in our sample. 
 We then extracted the source spectra (with extraction radius in the range $\simeq 25\arcsec-50\arcsec$) by applying the filter expressions $\#XMMEA\_EM$ (for MOS) and
$\#XMMEA\_EP$ (for pn) and added the expression $FLAG == 0$ in order to reject events close to CCD gaps or bad pixels. 
We also accounted for all the valid patterns (PATTERN
in [0:12]) for the two MOS cameras while restricted the analysis to single and double events (PATTERN in [0:4]) for pn.  
When possible, the related background spectra were extracted on an annulus surrounding the target source or on a circle close to it. 

The EPIC source (background-corrected) spectra were re-binned to have at 
least  25 counts per energy bin for bright source, while we relaxed this requirement and used 15 counts per 
bin for fainter targets. Further,  the spectra  (as well as  the ancillary files and response matrices) were 
imported and used in XSPEC (version 12.9.0) to manage the spectral analysis and for fitting purposes.

For each target we inspected the spectrum and fit it with a model consisting in an absorbed power-law (model {\it A} in Table \ref{tablexrayanalysis}) with hydrogen column density 
fixed to the value given (fifth column) by the {\it NH Tool}\footnote{The {\it NH Tool} is available at  {\tt http://www.nsatlas.org}.} (which is based on \citealt{kalberla2005}) 
towards the source coordinates. In a few cases, the fit procedure converged towards meaningless values of the power-law index, thus forcing us to fix it to the value $\Gamma=1.7$ 
(in accordance to default value used in the 3XMM catalogue) thus leaving the power-law normalization as the single free fitting parameter\footnote{We remind that in such cases 
(as investigated by \citealt{watson2009}) varying the shape of the power-law would result in a flux change up to a few percent.}. The XSPEC implementation of the model is {\it const*phabs*powerlaw},
 with the {\it const} accounting for cross-calibration issues among the {\it XMM}-Newton instruments. For some cases, we fixed both the hydrogen column density and the power-law index to 
the default values used in compiling the  3XMM catalogue, i.e.  $n_H=3\times 10^{20}$ cm$^{-1}$ and $\Gamma=1.7$.

{  When we noted the typical signatures of obscured sources (usually active galactic nuclei of the Seyfert 2 type with $n_H>10^{22}$, see e.g. \citealt{matt2002} for a review), 
we used a simplified version of the models described in \cite{lamassa2014}, i.e. {\it const*phabs*pcfabs*powerlaw} (within XSPEC), where 
{\it pcfabs} represents a partial covering fraction absorption depending on the photo-electric cross-section, { {\it $n^{int}_H$} the intrinsic hydrogen column density} and $f$ a covering  
factor ranging between 0 (free source) to 1 (a full spherical envelope surrounding the source). We verified that this simple model results in acceptable fits and 
in best fit parameters consistent with those found by using more sophisticated models.}

The results of this analysis are reported in Table \ref{tablexrayanalysis}. Here,
for any source of the GiX catalogue we give the absorbed (third column) and unabsorbed (fourth column) flux in the $0.2-12$ keV energy band (i.e., the {\it XMM}-Newton entire range) obtained by 
fitting the associated spectra within the XSPEC package with one of the models (second column) described above. When the flux estimate lacks of the ($90\%$ confidence level) errors, the corresponding number represents an upper-limit.  
{We also give the hydrogen column density $n_H$ (fifth column), the power-law index $\Gamma$ (sixth column), as well as the {  $n^{int}_H$} and $f$ values for the B model sources (last two columns).

In the latter case the $n^{int}_H$ estimates confirmed the idea of obscured AGN as the sources of the spectra.}

\section{Candidate massive black holes from the $X$-ray/Radio fundamental plane}
\label{correlation}
{We searched for radio counterparts by cross-matching our sample of $X$-ray nuclear sources (consistent with the galaxy center, within the positional errors) 
with the FIRST and NVSS catalogues at 1.4 GHz, as well as RG2007 at 1.4 GHz, 4.9 GHz and 8.4 GHz.
Consequently, we found 19 $X$-ray sources {  (characterized by an $X$-ray detection likelihood larger than $\simeq 16.6$)} 
correlating in position\footnote{  The distance between {  an $X$-ray source} and its radio counterpart was evaluated by using the well known haversine formula while the associated error was calculated 
by propagating correctly the uncertainties on both the celestial coordinates. In this respect, the error on the position of a radio source in the FIRST catalogue was derived by using the empyrical relation described in 
{\tt http://sundog.stsci.edu/first/catalogs/} {\tt readme$\_$13jun05.html$\#$evlacalibration} and, then, associated to the right ascension and declination, respectively.
For the NVSS sources, the error in position was simply read from the catalogue itself. In the case of RG2007, the positional error was obtained by summing in quadrature the 
respective uncertainties in the celestial coordinates.} with radio ones 
{  within $\simeq 2\arcsec$, apart from sources $\#$28 and $\#$33 which have radio counterparts at distance of 
$\simeq 4.2\arcsec$ (${\rm NVSS\,J125837+271033}$) and $\simeq 4.5\arcsec$ (${\rm NVSS\,J120222+295143}$), respectively. 
Since the positional errors (see Table \ref{tablemerloni}) on the separation between these two $X$-ray sources and their radio counterparts are $\simeq 4\arcsec$ and $\simeq 2\arcsec$, 
the association is less robust. Anyway, we maintain sources $\#$28 and $\#$33 in our final list of massive candidate black holes.}}

In Figure \ref{figfinal}, we give, for each source, the Sloan {\it r} 
band image (color inverted and in log scale) centered on the galaxy target. In each panel, the source position is
indicated by a red circle having radius of $3\arcsec$ centered on the $X$-ray source coordinates. 

{In the case of FIRST and NVSS, from the 1.4 GHz radio flux density of the identified sources, we can estimate the 5 GHz flux assuming a flat spectrum\footnote{  
We note that the obtained black hole mass values slightly depend on the index $\alpha_R$ characterizing the radio spectral energy distribution. In particular, the black hole mass 
scales with the factor $(1.4 GHz/5 GHz)^{1.28\alpha_R}$, get reducing by $\simeq 50\%$ for $\alpha_R=0.4$.}, i.e. $F(\nu) \propto \nu ^{-\alpha_R}$ with $\alpha_R=0$, while for source $\#1$ (Mrk 1303 also known as UM 444) the radio flux was directly taken from Table 2 in RG2007.}   
 
Then, after restricting the $X$-ray flux of our sources to the $2-10$ keV band,  we used the \citet{merloni2003} fundamental plane, i.e. 
\begin{eqnarray}
\log (M_{BH})\simeq 16.3+\log (D) \nonumber 
\\ +1.28(\log (F_{5~GHz}) -0.60\log (F_{2-10~keV}))\pm 1.06
\label{eqfundamentalplane}
\end{eqnarray}
in order to get an estimate of the mass of the candidate black holes.  In the previous relation, $D$ is the source distance expressed in Mpc and the last term
corresponds to the intrinsic scatter in the fundamental plane relation. {  In Table \ref{tablemerloni}, we give the main data corresponding to the {19} sources of interest, i.e. 
the ID used in the present paper, the galaxy distance (in Mpc) as derived from the NASA Sloan Atlas, the J2000 coordinates (in degrees) of the $X$-ray target and the positional error ($Err^X$),
the $X$-ray flux in the $2-10$ keV band, the coordinates of the radio counterpart along with the positional uncertainty ($Err^{radio}$), 
the 5 GHz flux, the $X$-ray and radio source separation and associated error, the mass of the black hole,
the mass of the galaxy $M_{gal}$ as derived using the SDSS data via the {\it kcorrect} code of \citet{blanton2007}, and the dynamical mass $M_{gal}^{dyn}$ obtained by using the 
mass estimator described in \cite{padmanabhan}}. In particular, the three-dimensional velocity $v$ of a test object orbiting a mass $M(r)$ at distance $r$ is simply given by
\begin{equation}
v^2(r)=\frac{GM(r)}{r}.
\end{equation}
To compute the dynamical mass, one must choose a characteristic radius and relate the observed dispersion velocity $\sigma$ at that radius to the above circular velocity. 
Here, we fixed $r$ to the Petrosian $50\%$ light radius ($R_{50}$ as derived from SDSS r band) and  assumed  $v^2(r)=\eta \sigma^2$, {  where $\eta$ is $2$ or $3$ in case of spirals or ellipticals, respectively.}

Following \cite{bernardi2003} (and references therein) we applied an empirical correction to the measured dispersion $\sigma$ to account for the fact that the dispersion velocity estimated by the SDSS spectra is not 
at the Petrosian $50\%$  radius but a the fiber diameter of $3\arcsec$. Then the corrected dispersion velocity is
\begin{equation}
\sigma_{corr}=\sigma \left(\frac{8r_{fiber}}{r_{eff}}\right)^{0.04},
\end{equation}
where $r_{fiber}\simeq 1.5\arcsec$ and $r_{eff}$ is the effective radius of the galaxy measured in arcseconds which we set again equal to $R_{50}$. Finally, our dynamical mass estimator reads out as 
\begin{equation}
M_{gal}^{dyn}=\frac{\eta\sigma_{corr}^2 R_{50}}{G}.
\end{equation}
{The above algorithm (except for $\# 40$ and $\# 51$) allows one to estimate the dynamical mass of the galaxy, but this value has not to be consider too robust as
other mass estimators could give different values. {In Table \ref{tablemerloni}, we always assumed $\eta=3$.}
In the last column of Table \ref{tablemerloni} we also give the {$X$-ray} accretion efficiency $\epsilon_X$ which represents a lower limit to the true accretion efficiency value, i.e.

\begin{equation}
\epsilon_X\simeq \frac{L_{0.2-12~keV}}{L_{Edd}},
\end{equation}
}
where $L_{Edd}\simeq 1.38\times 10 ^{38} \left(\frac{M_{BH}}{M_{\odot}}\right)$ erg s$^{-1}$. 

{In all the cases for which the $X$-ray target is found within a distance of $\simeq 3\arcsec$ from the optical centre, the corresponding source was considered to be a nuclear black hole candidate. When the distance
is greater than  $3\arcsec$, the source could not be a non-nuclear black hole 
candidate (possibly a ULX, see in this respect \citealt{soria2008} for a discussion on the number of ULXs found in low mass galaxies). Note also that for sources $\# 35$ (NGC 4117) and $\# 50$ (NGC 4395), our estimate of the candidate black hole mass is a factor $\simeq 10$ larger (but still consistent due to
the large intrinsic scatter of the fundamental plane relation) than the values derived by \citet{wau2002} and \citet{denbrok2015}, respectively, 
when studying the gas dynamics in the galaxies and/or using the virial assumption for the broad emission lines while determining the broad-line region size from either reverberation mapping or optical luminosity.}

{In Figure \ref{figmerloni}, we give the $X$-ray/radio fundamental plane (adapted from \citealt{merloni2003})
super-imposing our results (dodger blue empty stars). Here, we consider Galactic BHs (GBH), Liners-Transition and quasi stellar objects (L-T and QSO, respectively), 
Seyfert nuclei (Sy) and undefined sources.
As evident, our sample lies on the $X$-ray/radio fundamental plane so that we are justified in using eq. (\ref{eqfundamentalplane}) to infer the black hole mass.
This is also in accordance to \citealt{gultekin2014} who, by studying a sample of low-mass AGNs in the $X$-ray and radio bands, concluded that the fundamental 
plane is a good mass estimator and is suitable for searching for IMBH candidates  . In this respect, we also consider in the same figure 
the IMBH candidates (filled black circles, and mass less than $10^{6.3}$ M$_{\odot}$) corresponding to the eleven low-mass AGNs in \citealt{gultekin2014}. 
We also insert the observations corresponding to IMBHs (open circles) supposed to be hosted in selected globular clusters and nearby dwarf galaxies (\citealt{nucita2008,wrobel2011,webb2012,nyland2012,nucita2013a, nucita2013b,manni2015,mezcua2015,earnshaw2016a,earnshaw2016b}), 
whose associated mass can be determined via dynamical methods.}

\onecolumn
%  \addtocounter{table}{1}
%\begin{landscape}
 \renewcommand{\thefootnote}{\fnsymbol{footnote}}
 \renewcommand{\arraystretch}{0.81}
 \footnotesize
 \begin{longtable}{|c|c|c|c|c|c|c|c|c|c|c|}
 \caption{The GiX catalogue obtained by cross-correlating the NASA-Sloan Atlas and the 3XMM-DR5 databases. Columns are: our identifier, the {\it XMM}-Newton observation ID, the NASA-Sloan Atlas ID, the common target name, the (J2000) optical coordinates 
associated to the galaxies, the distance (in arcseconds) betwen the Sloan source and its $X$-ray counterpart, the galaxy mass, the quality flag associated to the $X$-ray source, {and its counts as derived by the 3XMM-DR5 catalogue}.
{  Source $\#$ 42 was also in the {\it XMM}-Newton field of view during the observation $0550960601$. However, since this particular observation was affected by large flares for most of the exposure window, we avoided to use it.}
}
 \label{tableobsidxmm}\\
 \hline\hline 

    \multicolumn{1}{c}{\textbf{Src \#}} &
    \multicolumn{1}{c}{\textbf{OBSID}} &
    \multicolumn{1}{c}{\textbf{NSAID}} &
    \multicolumn{1}{c}{\textbf{Name}} &
    \multicolumn{1}{c}{\textbf{RA (J2000)}} &
    \multicolumn{1}{c}{\textbf{DEC (J2000)}} &
    \multicolumn{1}{c}{\textbf{s}} &
    \multicolumn{1}{c}{\textbf{$M_{gal}$}} &
    \multicolumn{1}{c}{\textbf{SF}}&
    \multicolumn{1}{c}{\textbf{Counts}} \\
     \multicolumn{1}{c}{\textbf{}} &
    \multicolumn{1}{c}{\textbf{}} &
    \multicolumn{1}{c}{\textbf{}} &
    \multicolumn{1}{c}{\textbf{}} &
    \multicolumn{1}{c}{\textbf{$\degr$}} &
    \multicolumn{1}{c}{\textbf{$\degr$}} &
    \multicolumn{1}{c}{\textbf{$\arcsec$}} &
    \multicolumn{1}{c}{\textbf{$M_{\odot}$}} &
    \multicolumn{1}{c}{\textbf{}} &
    \multicolumn{1}{c}{\textbf{}} \\
 \hline\hline
 \endfirsthead
 \multicolumn{9}{c}{{\tablename} \thetable{} -- Continued}\\
 \hline\hline
   
    \multicolumn{1}{c}{\textbf{Src \#}} &
    \multicolumn{1}{c}{\textbf{OBSID}} &
    \multicolumn{1}{c}{\textbf{NSAID}} &
    \multicolumn{1}{c}{\textbf{Name}} &
    \multicolumn{1}{c}{\textbf{RA (J2000)}} &
    \multicolumn{1}{c}{\textbf{DEC (J2000)}} &
    \multicolumn{1}{c}{\textbf{s}} &
    \multicolumn{1}{c}{\textbf{$M_gal$}} &
    \multicolumn{1}{c}{\textbf{SF}} &
    \multicolumn{1}{c}{\textbf{Counts}}\\
    \multicolumn{1}{c}{\textbf{}} &
    \multicolumn{1}{c}{\textbf{}} &
    \multicolumn{1}{c}{\textbf{}} &
    \multicolumn{1}{c}{\textbf{}} &
    \multicolumn{1}{c}{\textbf{$\degr$}} &
    \multicolumn{1}{c}{\textbf{$\degr$}} &
    \multicolumn{1}{c}{\textbf{$\arcsec$}} &
    \multicolumn{1}{c}{\textbf{$M_{\odot}$}} &
    \multicolumn{1}{c}{\textbf{}}&
    \multicolumn{1}{c}{\textbf{}}\\
 \hline\hline 
 \endhead
 \multicolumn{9}{c}{{Continued on Next Page\ldots}} \\
 \endfoot
 \hline\hline
 \endlastfoot
 % \multirow{15}{*}{}
  1 &  $0303561801$ & $  1120$ & Mrk 1303                 &  $175.05513$ & $-0.41171$ & $0.6$ &  $  7.99\times 10^{+08}$ &  0 & $   42.8  \pm    8.7  $ \\
  2 &  $0124710501$ & $103903$ & Mrk 58                   &  $194.77209$ & $27.64444$ & $2.7$ &  $  8.08\times 10^{+09}$ &  0 & $  117.2  \pm   20.0  $ \\
  3 &  $0108860501$ & $ 84201$ & 2MASX J08193880+2103521  &  $124.91171$ & $21.06439$ & $2.1$ &  $  2.11\times 10^{+09}$ &  0 & $   42.3  \pm   10.0  $ \\
  4 &  $0041180801$ & $163589$ & 2MASX J13293557+1141515  &  $202.39807$ & $11.69777$ & $1.9$ &  $  9.48\times 10^{+09}$ &  0 & $   62.4  \pm   14.2  $ \\
  5 &  $0205910101$ & $ 88939$ & NVSS J130916+292202      &  $197.31704$ & $29.36766$ & $2.0$ &  $  7.31\times 10^{+09}$ &  0 & $  233.4  \pm   21.0  $ \\
  6 &  $0403150201$ & $103917$ & 2MASX J12581865+2718387  &  $194.57766$ & $27.31082$ & $1.0$ &  $  4.53\times 10^{+09}$ &  0 & $  158.7  \pm   20.8  $ \\
  7 &  $0551280101$ & $113915$ & ACO 1413                 &  $178.99342$ & $23.45862$ & $0.6$ &  $  4.66\times 10^{+09}$ &  0 & $  120.7  \pm   19.8  $ \\
  8 &  $0150010601$ & $ 60414$ & 2MASX J12093747+4219081  &  $182.40619$ & $42.31888$ & $1.8$ &  $  9.44\times 10^{+09}$ &  0 & $   29.6  \pm    8.8  $ \\
  9 &  $0200530401$ & $ 40521$ & ANT Galaxy               &  $164.21248$ & $ 6.90612$ & $2.4$ &  $  6.06\times 10^{+09}$ &  0 & $  128.0  \pm   15.1  $ \\
 10 &  $0402370101$ & $ 50679$ & 2XMMi J092720.4+362407   &  $141.83526$ & $36.40184$ & $1.1$ &  $  5.01\times 10^{+08}$ &  0 & $  160.3  \pm   17.9  $ \\
 11 &  $0404120101$ & $169920$ & IC 800                   &  $188.48607$ & $15.35484$ & $0.8$ &  $  7.47\times 10^{+09}$ &  0 & $  231.6  \pm   23.0  $ \\
 12 &  $0679381101$ & $ 37345$ & SDSS J112910.56+582309.0 &  $172.29392$ & $58.38587$ & $2.5$ &  $  3.10\times 10^{+09}$ &  0 & $   35.5  \pm    8.4  $ \\
 13 &  $0101640901$ & $ 97727$ & SDSS J153510.78+232409.4 &  $233.79493$ & $23.40266$ & $1.6$ &  $  1.82\times 10^{+09}$ &  0 & $   41.2  \pm    9.3  $ \\
 14 &  $0204400101$ & $162072$ & 2XMM J123519.9+393110    &  $188.83349$ & $39.51918$ & $2.3$ &  $  1.51\times 10^{+08}$ &  0 & $  207.4  \pm   18.1  $ \\
 15 &  $0652310701$ & $103933$ & Mrk 55                   &  $194.35521$ & $27.40457$ & $0.2$ &  $  6.07\times 10^{+09}$ &  0 & $   60.7  \pm   12.0  $ \\
 16 &  $0204650301$ & $ 88791$ & SDSS J113939.20+315320.3 &  $174.91338$ & $31.88897$ & $1.9$ &  $  7.21\times 10^{+09}$ &  0 & $   26.1  \pm    7.0  $ \\
 17 &  $0124710401$ & $104152$ & 7W 1258+27W06            &  $195.14032$ & $27.63774$ & $2.6$ &  $  5.44\times 10^{+09}$ &  0 & $   58.2  \pm   13.6  $ \\
 18 &  $0504101701$ & $ 58897$ & 2MASX J10181928+3722419  &  $154.58017$ & $37.37842$ & $0.2$ &  $  7.63\times 10^{+09}$ &  0 & $  225.8  \pm   17.7  $ \\
 19 &  $0400570101$ & $ 15235$ & 2XMMi J144012.6+024744   &  $220.05294$ & $ 2.79542$ & $0.7$ &  $  2.87\times 10^{+09}$ &  0 & $  377.7  \pm   23.4  $ \\
 20 &  $0150010601$ & $ 60412$ & SDSS J120900.89+422830.9 &  $182.25372$ & $42.47526$ & $2.1$ &  $  3.42\times 10^{+09}$ &  0 & $   77.9  \pm   12.2  $ \\
 21 &  $0112270601$ & $118256$ & 2MASX J12035600+2025499  &  $180.98330$ & $20.43044$ & $1.1$ &  $  1.05\times 10^{+10}$ &  0 & $   27.7  \pm   7.2   $ \\
 22 &  $0505210601$ & $ 99202$ & Mrk 695                  &  $240.71243$ & $15.96107$ & $2.0$ &  $  4.00\times 10^{+09}$ &  0 & $   95.8  \pm   16.6  $ \\
 23 &  $0094383201$ & $162694$ & Mrk 447                  &  $194.54165$ & $24.34891$ & $3.0$ &  $  4.20\times 10^{+09}$ &  0 & $   18.4  \pm    5.7  $ \\
 24 &  $0109462201$ & $ 42891$ & 2MASX J13070847+5357446  &  $196.78517$ & $53.96242$ & $1.0$ &  $  1.76\times 10^{+09}$ &  0 & $   37.8  \pm    8.1  $ \\
 25 &  $0103260801$ & $157523$ & FIRST J095310.3+075224   &  $148.29327$ & $ 7.87361$ & $1.0$ &  $  8.39\times 10^{+09}$ &  0 & $  256.1  \pm   19.3  $ \\
 26 &  $0606030101$ & $ 37086$ & MCG+10-16-025            &  $163.86892$ & $57.90620$ & $2.0$ &  $  7.47\times 10^{+09}$ &  0 & $  105.0  \pm   16.7  $ \\
 27 &  $0025540301$ & $ 84441$ & SDSS J083749.83+254804.8 &  $129.45764$ & $25.80136$ & $0.8$ &  $  5.83\times 10^{+09}$ &  0 & $   73.4  \pm   11.1  $ \\
 28 &  $0403150201$ & $103591$ & NVSS J125837+271033      &  $194.65534$ & $27.17655$ & $1.8$ &  $  5.25\times 10^{+09}$ &  0 & $   81.2  \pm   16.8  $ \\
 29 &  $0504101501$ & $19510$ & 2MASX J13463217+6423247  &  $206.63399$ & $64.39041$ & $0.6$ &   $  8.89\times 10^{+09}$ &  0 & $  168.9  \pm   17.3  $ \\
 30 &  $0202730101$ & $108428$ & SDSS J102217.95+212642.8 &  $155.57480$ & $21.44524$ & $1.2$ &  $  5.94\times 10^{+09}$ &  0 & $  704.8  \pm   30.9  $ \\
 31 &  $0674810701$ & $158462$ & NGC 3259                 &  $158.14524$ & $65.04115$ & $0.8$ &  $  8.92\times 10^{+09}$ &  0 & $  274.8  \pm   21.3  $ \\
 32 &  $0674810601$ & $ 63442$ & 2MASS J16315959+2437403  &  $247.99833$ & $24.62786$ & $0.2$ &  $  6.79\times 10^{+09}$ &  0 & $ 1105.3  \pm   36.9  $ \\
 33 &  $0555060301$ & $102563$ & NVSS J120222+295143      &  $180.59389$ & $29.86176$ & $2.9$ &  $  1.01\times 10^{+10}$ &  0 & $   33.4  \pm    8.3  $ \\
 34 &  $0124710801$ & $104185$ & 2XMM J130200.1+274657    &  $195.50061$ & $27.78271$ & $0.8$ &  $  9.35\times 10^{+09}$ &  0 & $ 3808.8  \pm   65.8  $ \\
 35 &  $0655800501$ & $ 60658$ & NGC 4117                 &  $181.94215$ & $43.12635$ & $0.2$ &  $  3.91\times 10^{+09}$ &  0 & $  326.3  \pm   21.8  $ \\
 36 &  $0204651201$ & $ 41675$ & NGC 3982                 &  $179.11726$ & $55.12528$ & $1.8$ &  $  1.00\times 10^{+10}$ &  0 & $ 1787.0  \pm   50.4  $ \\
 37 &  $0502211401$ & $103006$ & 2MASX J12221536+2821231  &  $185.56407$ & $28.35661$ & $0.3$ &  $  5.44\times 10^{+09}$ &  0 & $   88.4  \pm   11.9  $ \\
 38 &  $0504102001$ & $156668$ & 2MASX J08244333+2959238  &  $126.18034$ & $29.98987$ & $0.1$ &  $  9.41\times 10^{+09}$ &  0 & $ 4081.4  \pm   68.7  $ \\
 39 &  $0504100201$ & $ 74152$ & 2MASX J09591475+1259161  &  $149.81153$ & $12.98789$ & $0.4$ &  $  4.68\times 10^{+09}$ &  0 & $ 7322.1  \pm   90.6  $ \\
 40 &  $0651100401$ & $ 45387$ & Mrk 477                  &  $220.15872$ & $53.50453$ & $0.3$ &  $  1.00\times 10^{+10}$ &  0 & $ 3204.3  \pm   60.4  $ \\
 \hline                                                                                                                                                   
 \hline                                                                                                                                                   
 41 &  $0400570201$ & $ 27397$ & UGC 6192                 &  $167.30166$ & $61.39641$ & $1.4$ &  $ 8.21\times 10^{+08} $ &  1 & $   61.9  \pm  11.9   $ \\
 42 &  $0067340601$ & $ 73400$ &    ---                   &  $241.71258$ & $ 8.15796$ & $0.9$ &  $ 1.50\times 10^{+08} $ &  1 & $  164.5  \pm  24.6   $ \\
 43 &  $0404410101$ & $ 62285$ &    ---                   &  $ 18.85011$ & $ 0.63552$ & $2.0$ &  $ 1.82\times 10^{+08} $ &  1 & $  124.5  \pm  23.8   $ \\
 44 &  $0210280101$ & $ 84591$ & 2XMM J085735.4+274607    &  $134.39726$ & $27.76813$ & $2.6$ &  $ 2.88\times 10^{+08} $ &  1 & $  512.3  \pm  29.6   $ \\
 45 &  $0147210301$ & $ 99052$ & 2XMM J160531.8+174825    &  $241.38272$ & $17.80726$ & $0.5$ &  $ 1.74\times 10^{+09} $ &  1 & $  136.2  \pm  15.5   $ \\
 46 &  $0111260201$ & $165958$ & NGC 5879                 &  $227.44484$ & $57.00024$ & $1.5$ &  $ 8.14\times 10^{+09} $ &  1 & $  352.4  \pm  23.3   $ \\
 47 &  $0303550901$ & $ 10045$ & 2XMMi J082912.8+500652   &  $127.30277$ & $50.11465$ & $0.4$ &  $ 4.12\times 10^{+09} $ &  1 & $ 24289.5 \pm 172.5   $ \\
 48 &  $0674811101$ & $ 52675$ & 2MASX J12234282+5814459  &  $185.92846$ & $58.24623$ & $0.2$ &  $ 2.92\times 10^{+09} $ &  1 & $ 7622.3  \pm  92.8   $ \\
 49 &  $0149010201$ & $ 47963$ & IC 2461                  &  $139.99178$ & $37.19125$ & $0.8$ &  $ 8.13\times 10^{+09} $ &  1 & $19663.3  \pm 149.9   $ \\
 50 &  $0112521901$ & $ 89394$ & NGC 4395                 &  $186.45362$ & $33.54687$ & $0.1$ &  $ 1.27\times 10^{+09} $ &  1 & $12175.3  \pm 116.2   $ \\
 51 &  $0051760101$ & $ 14301$ & 2XMM J124635.3+022209    &  $191.64688$ & $ 2.36911$ & $0.1$ &  $ 6.86\times 10^{+09} $ &  1 & $82587.9  \pm 296.8   $
 \end{longtable}
 \normalsize
 \renewcommand{\thefootnote}{\arabic{footnote}}
\renewcommand{\arraystretch}{1.0}
%\end{landscape}
\twocolumn

{
\onecolumn
%  \addtocounter{table}{1}
%\begin{landscape}
 \renewcommand{\thefootnote}{\fnsymbol{footnote}}
 \renewcommand{\arraystretch}{0.81}
  \footnotesize
 \begin{longtable}{|c|c|c|c|c|c|c|c|}
 \caption{The absorbed (third column) and unabsorbed (fourth column) flux in the $0.2-12$ keV energy band for each of the GiX catalogue entries. The $X$-ray spectra were fitted within the XSPEC 
package using the model indicated in the second column (see text for details) and the values of the hydrogen column density, power-law index, 
{  intrinsic hydrogen column density} and covering fraction factor shown in the last four columns, respectively. {  For source $\#$ 51 (labeled with an asterisk), considering a single power-law resulted in residuals
at low energies. The residuals disappeared when considering a black-body component with temperature of {\it kT} $\simeq 0.14$ keV.}}
  \label{tablexrayanalysis}\\
 \hline\hline 
    \multicolumn{1}{c}{\textbf{Src \#}} &
    \multicolumn{1}{c}{\textbf{Model}} &
    \multicolumn{1}{c}{\textbf{F$_{0.2-12 keV}^{Abs}$}} &
    \multicolumn{1}{c}{\textbf{F$_{0.2-12 keV}^{UnAbs}$}} &
    \multicolumn{1}{c}{\textbf{$n_H$}} &
    \multicolumn{1}{c}{\textbf{$\Gamma$}} &
    \multicolumn{1}{c}{\textbf{$n_H^{int}$}} &
    \multicolumn{1}{c}{\textbf{$f$}} \\
    \multicolumn{1}{c}{\textbf{}} &
    \multicolumn{1}{c}{\textbf{}} &
    \multicolumn{1}{c}{\textbf{erg s$^{-1}$ cm$^{-2}$}} &
    \multicolumn{1}{c}{\textbf{erg s$^{-1}$ cm$^{-2}$}} &
    \multicolumn{1}{c}{\textbf{$10^{20}$ cm$^{-2}$}} &
    \multicolumn{1}{c}{\textbf{}} &
    \multicolumn{1}{c}{\textbf{$10^{22}$ cm$^{-2}$}} &
    \multicolumn{1}{c}{\textbf{}}    \\
 \hline\hline
 \endfirsthead
 \multicolumn{8}{c}{{\tablename} \thetable{} -- Continued}\\
 \hline\hline
    \multicolumn{1}{c}{\textbf{Src \#}} &
    \multicolumn{1}{c}{\textbf{Model}} &
    \multicolumn{1}{c}{\textbf{F$_{0.2-12 keV}^{Abs}$}} &
    \multicolumn{1}{c}{\textbf{F$_{0.2-12 keV}^{UnAbs}$}} &
    \multicolumn{1}{c}{\textbf{$n_H$}} &
    \multicolumn{1}{c}{\textbf{$\Gamma$}} &
    \multicolumn{1}{c}{\textbf{$n_H^{int}$}} &
    \multicolumn{1}{c}{\textbf{$f$}} \\
    \multicolumn{1}{c}{\textbf{}} &
    \multicolumn{1}{c}{\textbf{}} &
    \multicolumn{1}{c}{\textbf{erg s$^{-1}$ cm$^{-2}$}} &
    \multicolumn{1}{c}{\textbf{erg s$^{-1}$ cm$^{-2}$}} &
    \multicolumn{1}{c}{\textbf{$10^{20}$ cm$^{-2}$}} &
    \multicolumn{1}{c}{\textbf{}} &
    \multicolumn{1}{c}{\textbf{$10^{22}$ cm$^{-2}$}} &
    \multicolumn{1}{c}{\textbf{}}    \\
 \hline\hline                                                                                   
 \endhead
 \multicolumn{8}{c}{{Continued on Next Page\ldots}} \\
 \endfoot
 \hline\hline
 \endlastfoot
 % \multirow{15}{*}{}
  1 &  A   & $ 2.15^{+1.2}_{-1.2}  \times 10^{-14 }$ & $ 2.3 \times 10^{-14 }$  & $ 2.4$  & $ 1.7                  $ & $                           $ & $                         $ \\      
  2 &  A   & $ 1.7^{+2.5}_{ -0.8}  \times 10^{-14 }$ & $ 1.9 \times 10^{-14 }$  & $ 0.9$  & $ 2.28^{+0.96}_{-0.48} $ & $                           $ & $                         $ \\         
  3 &  A   & $ 2.8^{ +1.8}_{ -1.7} \times 10^{-14 }$ & $ 3.2 \times 10^{-14 }$  & $ 4.3$  & $ 1.7                  $ & $                           $ & $                         $ \\
  4 &  A   & $ 1.2^{ +0.7}_{-0.7}  \times 10^{-14 }$ & $ 1.3 \times 10^{-14 }$  & $ 3.0$  & $ 1.7                  $ & $                           $ & $                         $ \\
  5 &  A   & $ 1.2^{ +0.4}_{ -0.5} \times 10^{-14 }$ & $ 1.2 \times 10^{-14 }$  & $ 1.0$  & $ 1.7                  $ & $                           $ & $                         $ \\          
  6 &  A   & $ 1.4^{+0.5}_{-0.5}   \times 10^{-14 }$ & $ 1.6 \times 10^{-14 }$  & $ 3.0$  & $ 1.7                  $ & $                           $ & $                         $ \\
  7 &  A   & $ 1.1^{ +0.7}_{ -0.7} \times 10^{-14 }$ & $ 1.2 \times 10^{-14 }$  & $ 2.1$  & $ 1.7                  $ & $                           $ & $                         $ \\          
  8 &  A   & $ 2.0^{ +2.1}_{ -1.9} \times 10^{-14 }$ & $ 2.2 \times 10^{-14 }$  & $ 1.7$  & $ 1.7                  $ & $                           $ & $                         $ \\
  9 &  A   & $ 2.4^{ +1.7}_{ -0.6} \times 10^{-14 }$ & $ 3.3 \times 10^{-14 }$  & $ 3.0$  & $ 2.41^{+0.82}_{-0.61} $ & $                           $ & $                         $ \\          
 10 &  A   & $ 2.9^{ +1.5}_{ -1.0} \times 10^{-14 }$ & $ 2.9 \times 10^{-14 }$  & $ 1.4$  & $ 1.18^{+0.40}_{-0.38} $ & $                           $ & $                         $ \\           
 11 &  A   & $ 1.7^{ +0.3}_{-0.3}  \times 10^{-14 }$ & $ 1.8 \times 10^{-14 }$  & $ 3.0$  & $ 1.7                  $ & $                           $ & $                         $ \\                                                           
 12 &  A   & $ 2.7^{ +1.5}_{ -1.6} \times 10^{-14 }$ & $ 2.8 \times 10^{-14 }$  & $ 1.0$  & $ 1.7                  $ & $                           $ & $                         $ \\
 13 &  A   & $ 1.1^{ +0.6}_{ -0.5} \times 10^{-14 }$ & $ 1.2 \times 10^{-14 }$  & $ 4.3$  & $ 1.7                  $ & $                           $ & $                         $ \\
 14 &  A   & $ 2.9^{ +1.2}_{ -0.6} \times 10^{-14 }$ & $ 3.1 \times 10^{-14 }$  & $ 1.5$  & $ 1.79^{+0.45}_{-0.36} $ & $                           $ & $                         $ \\
 15 &  A   & $ 6.2^{ +2.6}_{ -2.8} \times 10^{-14 }$ & $ 6.5 \times 10^{-14 }$  & $ 0.9$  & $ 1.7                  $ & $                           $ & $                         $ \\                  
 16 &  A   & $ 3.2^{ +7.6}_{ -1.9} \times 10^{-14 }$ & $ 3.7 \times 10^{-14 }$  & $ 2.0$  & $ 1.82^{+1.54}_{-0.87} $ & $                           $ & $                         $ \\    
 17 &  A   & $ 4.3                 \times 10^{-14 }$ & $ 4.4 \times 10^{-14 }$  & $ 0.9$  & $ 1.7                  $ & $                           $ & $                         $ \\                
 18 &  A   & $ 4.2^{ +0.7}_{ -0.7} \times 10^{-14 }$ & $ 5.3 \times 10^{-14 }$  & $ 1.3$  & $ 2.83^{+0.27}_{-0.24} $ & $                           $ & $                         $ \\      
 19 &  A   & $ 2.8^{ +0.5}_{ -0.5} \times 10^{-14 }$ & $ 3.1 \times 10^{-14 }$  & $ 2.9$  & $ 1.7                  $ & $                           $ & $                         $ \\       
 20 &  A   & $ 3.2^{ +1.2}_{-1.2}  \times 10^{-14 }$ & $ 3.5 \times 10^{-14 }$  & $ 3.0$  & $ 1.7                  $ & $                           $ & $                         $ \\                                              
 21 &  A   & $ 3.3^{ +2.8}_{-2.8}  \times 10^{-14 }$ & $ 3.6 \times 10^{-14 }$  & $ 3.0$  & $ 1.7                  $ & $                           $ & $                         $ \\                                 
 22 &  A   & $ 4.5^{ +1.9}_{ -2.1} \times 10^{-14 }$ & $ 5.0 \times 10^{-14 }$  & $ 3.4$  & $ 1.7                  $ & $                           $ & $                         $ \\                
 23 &  A   & $ 3.5^{ +2.4}_{-2.4}  \times 10^{-14 }$ & $ 3.8 \times 10^{-14 }$  & $ 3.0$  & $ 1.7                  $ & $                           $ & $                         $ \\                                                                                    
 24 &  A   & $ 3.6^{ +1.8}_{-1.8}  \times 10^{-14 }$ & $ 4.0 \times 10^{-14 }$  & $ 3.0$  & $ 1.7                  $ & $                           $ & $                         $ \\                                       
 25 &  A   & $ 2.8^{ +0.7}_{-0.3}  \times 10^{-14 }$ & $ 3.8 \times 10^{-14 }$  & $ 3.0$  & $ 2.38^{+0.63}_{-0.47} $ & $                           $ & $                         $ \\                                                      
 26 &  A   & $ 5.0^{ +3.4}_{ -1.7} \times 10^{-14 }$ & $ 5.1 \times 10^{-14 }$  & $ 0.6$  & $ 1.44^{+0.47}_{-0.40} $ & $                           $ & $                         $ \\                                              
 27 &  A   & $ 6.3^{ +5.8}_{ -2.5} \times 10^{-14 }$ & $ 6.6 \times 10^{-14 }$  & $ 3.6$  & $ 1.29^{+0.56}_{-0.55} $ & $                           $ & $                         $ \\                                
 28 &  A   & $ 5.4^{ +3.1}_{-3.1}  \times 10^{-14 }$ & $ 6.0 \times 10^{-14 }$  & $ 3.0$  & $ 1.7                  $ & $                           $ & $                         $ \\                                                 
 29 &  A   & $ 2.3^{ +1.7}_{-0.8}  \times 10^{-14 }$ & $ 2.5 \times 10^{-14 }$  & $ 2.2$  & $ 1.48^{+0.78}_{-0.83} $ & $                           $ & $                         $ \\
 30 &  A   & $ 8.1^{ +1.2}_{-1.0}  \times 10^{-14 }$ & $ 8.6 \times 10^{-14 }$  & $ 2.0$  & $ 1.59^{+0.12}_{-0.12} $ & $                           $ & $                         $ \\
 31 &  B   & $ 1.4                 \times 10^{-13 }$ & $ 1.5 \times 10^{-13 }$  & $ 1.7$  & $ 2.19^{+0.18}_{-0.23} $ & $ 18.36^{+7.37}_{-5.71}   $ & $ 0.98^{+0.01}_{-0.1}    $ \\
 32 &  B   & $ 1.4^{ +0.2}_{-0.7}  \times 10^{-13 }$ & $ 1.9 \times 10^{-13 }$  & $ 3.8$  & $ 2.35^{+0.13}_{-0.13} $ & $ 18.52^{+75.15}_{-12.31}   $ & $ 0.70^{+0.23}_{-0.30}    $ \\
 33 &  A   & $ 9.2                 \times 10^{-14 }$ & $ 9.7 \times 10^{-14 }$  & $ 1.6$  & $ 1.7                  $ & $                           $ & $                         $ \\
 34 &  A   & $ 4.3^{ +0.1}_{-0.1}  \times 10^{-13 }$ & $ 6.0 \times 10^{-13 }$  & $ 0.9$  & $ 1.7                  $ & $                           $ & $                         $ \\
 35 &  B   & $ 4.3^{ +1,1}_{-1.8}  \times 10^{-13 }$ & $ 4,6 \times 10^{-13 }$  & $ 1.4$  & $ 1.7 $ & $ 51.58^{+13.91}_{-12.21}   $ & $ 0.98^{+0.01}_{-0.02} $ \\
 36 &  B   & $ 3.5^{ +2.5}_{ -3.0} \times 10^{-13 }$ & $ 4.2 \times 10^{-13 }$  & $ 1.2$  & $ 3.08^{+0.31}_{-0.35} $ & $ 21.23^{+12.11}_{-6.77}    $ & $ 0.985^{+0.012}_{-0.050} $ \\
 37 &  A   & $ 1.8^{ +0.6}_{ -0.7} \times 10^{-13 }$ & $ 1.9 \times 10^{-13 }$  & $ 1.8$  & $ 1.7                  $ & $                           $ & $                         $ \\
 38 &  B   & $ 1.4^{ +0.1}_{ -0.2} \times 10^{-12 }$ & $ 1.5 \times 10^{-12 }$  & $ 3.8$  & $ 1.30^{+0.06}_{-0.15} $ & $ 17.71^{+1.56}_{-1.67}     $ & $ 0.992^{+0.006}_{-0.042} $ \\
 39 &  B   & $ 1.3^{ +0.1}_{ -0.1} \times 10^{-12 }$ & $ 1.3 \times 10^{-12 }$  & $ 3.2$  & $ 1.90^{+0.05}_{-0.12} $ & $ 0.92^{+0.08}_{-0.09}      $ & $ > 0.945 $ \\
 40 &  B   & $ 2.3^{ +0.2}_{ -0.2} \times 10^{-12 }$ & $ 2.3 \times 10^{-12 }$  & $ 1.3$  & $ 1.75^{+0.05}_{-0.10} $ & $ 25.84^{+3.03}_{-2.20}     $ & $ 0.990^{+0.008}_{-0.040} $ \\
 41 &  A   & $ 4.2^{ +3.6}_{ -3.8} \times 10^{-15 }$ & $ 4.4 \times 10^{-15 }$  & $ 0.7$  & $ 1.7                  $ & $                           $ & $                         $ \\
 42 &  A   & $4.9^{ +0.5}_{ -2.9}  \times 10^{-14 }$ & $ 5.0 \times 10^{-14 }$  & $ 4.0$  & $ 1.04^{+0.73}_{-0.63} $ & $                           $ & $                         $ \\
 43 &  A   & $ 5.5^{ +5.7}_{ -2.8} \times 10^{-14 }$ & $ 5.8 \times 10^{-14 }$  & $ 3.2$  & $ 1.37^{+1.17}_{-0.85} $ & $                           $ & $                         $ \\
 44 &  A   & $ 2.6^{ +0.6}_{ -0.5} \times 10^{-14 }$ & $ 2.8 \times 10^{-14 }$  & $ 3.1$  & $ 1.70^{+0.17}_{-0.17} $ & $                           $ & $                         $ \\
 45 &  A   & $ 3.5^{ +13.9}_{ -2.2}\times 10^{-14 }$ & $ 3.8 \times 10^{-14 }$  & $ 3.5$  & $ 1.54^{+1.46}_{-1.05} $ & $                           $ & $                         $ \\
 46 &  A   & $ 4.8^{ +1.1}_{ -0.7} \times 10^{-14 }$ & $ 5.5 \times 10^{-14 }$  & $ 1.5$  & $ 2.20^{+0.32}_{-0.35} $ & $                           $ & $                         $ \\
 47 &  A   & $ 2.06^{ +0.03}_{ -0.04}\times 10^{-12 }$ & $ 3.2\times 10^{-12 }$ & $ 4.0$  & $ 2.61^{+0.02}_{-0.02} $ & $                           $ & $                         $ \\
 48 &  A   & $ 2.9^{ +0.1}_{ -0.1} \times 10^{-12 }$ & $ 3.1 \times 10^{-12 }$  & $ 1.2$  & $ 1.83^{+0.03}_{-0.03} $ & $                           $ & $                         $ \\
 49 &  B   & $ 5.4^{ +0.1}_{ -0.2} \times 10^{-12 }$ & $ 5.4 \times 10^{-12 }$  & $ 1.6$  & $ 0.60^{+0.03}_{-0.05} $ & $ 4.30^{+0.22}_{-0.05}      $ & $ 0.995^{+0.005}_{-0.005} $ \\
 50 &  B   & $ 8.4^{ +0.2}_{ -0.3} \times 10^{-12 }$ & $ 8.4 \times 10^{-12 }$  & $ 1.3$  & $ 0.91^{+0.03}_{-0.05} $ & $ 4.81^{+0.32}_{-0.29}      $ & $ 0.975^{+0.015}_{-0.025} $ \\
 51*&  A   & $ 1.28^{ +0.02}_{ -0.03}\times 10^{-11 }$ & $ 1.8 \times 10^{-11 }$& $ 1.8$  & $ 2.91^{+0.03}_{-0.03} $ & $                           $ & $                         $ 
 \end{longtable}                                                                                                      
 \normalsize                                                                                                          
 \renewcommand{\thefootnote}{\arabic{footnote}}                                                                       
\renewcommand{\arraystretch}{1.0}                                                                                     
%\end{landscape}                                                                                                       
\twocolumn                                                                                                            
}

{
\onecolumn                                                                                                            
%  \addtocounter{table}{1}                                                                                            
 \begin{landscape}
 \renewcommand{\thefootnote}{\fnsymbol{footnote}}
 \renewcommand{\arraystretch}{0.81}
  \footnotesize
 \begin{longtable}{|c|c|c|c|c|c|c|c|c|c|c|c|c|c|c|}
 \caption{{  A sub-sample of the GiX sources correlating in position with the 1.4 GHz sources of the {  FIRST, NVSS and RG2007 catalogues}. For these sources we give the galaxy distance, the $X$-ray position and associated error,  
the unabsorbed flux in $2-10$ keV band, the radio counterpart position and error, the 5 GHz flux, the separation (and associated uncertainty) between the $X$-ray and the radio source, 
the estimated central black hole mass, the mass of the galaxy as derived by the SDSS observations, the dynamical 
mass and the black hole $X$-ray accretion efficiency $\epsilon_X$ (see text for details). Note that the error associated to the black hole mass is $1.06$ dex as derived from the fundamental plane relation.}}
 \label{tablemerloni}\\
 \hline\hline 
    \multicolumn{1}{c}{\textbf{Src \#}} &
    \multicolumn{1}{c}{\textbf{D}} &
    \multicolumn{1}{c}{\textbf{RA$^X$ (J2000)}} &
    \multicolumn{1}{c}{\textbf{DEC$^X$ (J2000)}} &
    \multicolumn{1}{c}{\textbf{$Err^X$}} &
    \multicolumn{1}{c}{\textbf{$F_{2-10 keV}$}} &
    \multicolumn{1}{c}{\textbf{RA$^{radio}$ (J2000)}} &
    \multicolumn{1}{c}{\textbf{DEC$^{radio}$ (J2000)}} &
    \multicolumn{1}{c}{\textbf{$Err^{radio}$}} &
    \multicolumn{1}{c}{\textbf{$F_{5 GHz}$}} &
    \multicolumn{1}{c}{\textbf{s}} &
    \multicolumn{1}{c}{\textbf{log $M_{BH}$}} &
    \multicolumn{1}{c}{\textbf{log $M_{gal}$}} &
    \multicolumn{1}{c}{\textbf{log $M_{gal}^{dyn}$}} &
    \multicolumn{1}{c}{\textbf{$\epsilon_X$}} \\
    \multicolumn{1}{c}{\textbf{ }} &
    \multicolumn{1}{c}{\textbf{Mpc}} &
    \multicolumn{1}{c}{\textbf{$\degr$}} &
    \multicolumn{1}{c}{\textbf{$\degr$}} &
    \multicolumn{1}{c}{\textbf{$\arcsec$}} &
    \multicolumn{1}{c}{\textbf{erg s$^{-1}$ cm$^{-2}$}} &
    \multicolumn{1}{c}{\textbf{$\degr$}} &
    \multicolumn{1}{c}{\textbf{$\degr$}} &
    \multicolumn{1}{c}{\textbf{$\arcsec$}} &
    \multicolumn{1}{c}{\textbf{erg s$^{-1}$ cm$^{-2}$}} &
    \multicolumn{1}{c}{\textbf{$\arcsec$}} &
    \multicolumn{1}{c}{\textbf{$M_{\odot}$}} &
    \multicolumn{1}{c}{\textbf{$M_{\odot}$}} &
    \multicolumn{1}{c}{\textbf{$M_{\odot}$}} &
    \multicolumn{1}{c}{\textbf{$ $}}\\
 \hline\hline
 \endfirsthead
 \multicolumn{15}{c}{{\tablename} \thetable{} -- Continued}\\
 \hline\hline
    \multicolumn{1}{c}{\textbf{Src \#}} &
    \multicolumn{1}{c}{\textbf{D}} &
    \multicolumn{1}{c}{\textbf{RA$^X$ (J2000)}} &
    \multicolumn{1}{c}{\textbf{DEC$^X$ (J2000)}} &
    \multicolumn{1}{c}{\textbf{$Err^X$}} &
    \multicolumn{1}{c}{\textbf{$F_{2-10 keV}$}} &
    \multicolumn{1}{c}{\textbf{RA$^{radio}$ (J2000)}} &
    \multicolumn{1}{c}{\textbf{DEC$^{radio}$ (J2000)}} &
    \multicolumn{1}{c}{\textbf{$Err^{radio}$}} &
    \multicolumn{1}{c}{\textbf{$F_{5 GHz}$}} &
    \multicolumn{1}{c}{\textbf{s}} &
    \multicolumn{1}{c}{\textbf{log $M_{BH}$}} &
    \multicolumn{1}{c}{\textbf{log $M_{gal}$}} &
    \multicolumn{1}{c}{\textbf{log $M_{gal}^{dyn}$}} &
    \multicolumn{1}{c}{\textbf{$\epsilon_X$}} \\
    \multicolumn{1}{c}{\textbf{ }} &
    \multicolumn{1}{c}{\textbf{Mpc}} &
    \multicolumn{1}{c}{\textbf{$\degr$}} &
    \multicolumn{1}{c}{\textbf{$\degr$}} &
    \multicolumn{1}{c}{\textbf{$\arcsec$}} &
    \multicolumn{1}{c}{\textbf{erg s$^{-1}$ cm$^{-2}$}} &
    \multicolumn{1}{c}{\textbf{$\degr$}} &
    \multicolumn{1}{c}{\textbf{$\degr$}} &
    \multicolumn{1}{c}{\textbf{$\arcsec$}} &
    \multicolumn{1}{c}{\textbf{erg s$^{-1}$ cm$^{-2}$}} &
    \multicolumn{1}{c}{\textbf{$\arcsec$}} &
    \multicolumn{1}{c}{\textbf{$M_{\odot}$}} &
    \multicolumn{1}{c}{\textbf{$M_{\odot}$}} &
    \multicolumn{1}{c}{\textbf{$M_{\odot}$}} &
    \multicolumn{1}{c}{\textbf{$ $}}\\
 \hline\hline 
 \endhead
 \multicolumn{15}{c}{{Continued on Next Page\ldots}} \\
 \endfoot
 \hline\hline
 \endlastfoot
 % \multirow{15}{*}{}
  1 &     $90.4  $   & $175.05527$  & $-0.41180$ & $ 1.7 $ &     $  1.2\times 10^{-14} $ & $175.05500 $&  $-0.41167 $&   $1.8 $&    $ 7.1\times 10^{-17} $  & $ 1.1 \pm  2.3  $ &   $ 8.28   $ &  $ 8.90 $  & $ 9.11 $  &  $  8.6 \times 10^{-7} $ \\
  2 &     $74.4  $   & $194.77145$  & $27.64394$ & $ 1.6 $ &     $  5.1\times 10^{-15} $ & $194.77195 $&  $27.64426 $&   $1.2 $&    $ 2.1\times 10^{-17} $  & $ 2.0 \pm  1.7  $ &   $ 7.80   $ &  $ 9.91 $  & $ 8.72 $  &  $  1.4 \times 10^{-6} $ \\
  5 &     $86.0  $   & $197.31680$  & $29.36716$ & $ 0.9 $ &     $  6.3\times 10^{-15} $ & $197.31708 $&  $29.36746 $&   $1.0 $&    $ 2.3\times 10^{-17} $  & $ 1.4 \pm  1.0  $ &   $ 7.84   $ &  $ 9.86 $  & $ 9.21 $  &  $  1.1\times 10^{-6}  $ \\   
  6 &     $101.5  $   & $194.57782$  & $27.31107$ & $ 1.2 $ &     $  8.0\times 10^{-15} $ & $194.57749 $&  $27.31093 $&   $0.6 $&    $ 2.5\times 10^{-17} $  & $ 1.2 \pm  1.2  $ &  $ 7.87   $ &  $ 9.66 $  & $ 9.65 $  &  $  1.9\times 10^{-6}  $ \\   
 15 &     $66.4  $   & $194.35523$  & $27.40452$ & $ 1.5 $ &     $  3.3\times 10^{-14} $ & $194.35525 $&  $27.40492 $&   $1.1 $&    $ 1.5\times 10^{-17} $  & $ 1.4 \pm  1.5  $ &   $ 6.95   $ &  $ 9.78 $  & $ 9.93 $  &  $  2.8\times 10^{-5}  $ \\   
 22 &     $144.5  $   & $240.71251$  & $15.96162$ & $ 1.7 $ &     $  2.6\times 10^{-14} $ & $240.71273 $&  $15.96104 $&   $1.1 $&    $ 1.8\times 10^{-17} $  & $ 2.2 \pm  1.7  $ &  $ 7.46   $ &  $ 9.60 $  & $ 9.42 $  &  $  3.1\times 10^{-5}  $ \\   
 24 &     $120.7  $   & $196.78472$  & $53.96239$ & $ 1.8 $ &     $  2.1\times 10^{-14} $ & $196.78549 $&  $53.96223 $&   $0.9 $&    $ 2.1\times 10^{-17} $  & $ 1.7 \pm  1.2  $ &  $ 7.52   $ &  $ 9.24 $  & $ 9.50 $  &  $  1.5\times 10^{-5}  $ \\   
 25 &     $167.1  $   & $148.29339$  & $ 7.87334$ & $ 0.7 $ &     $  9.1\times 10^{-15} $ & $148.29320 $&  $ 7.87349 $&   $0.8 $&    $ 3.7\times 10^{-17} $  & $ 0.9 \pm  0.9  $ &  $ 8.28   $ &  $ 9.92 $  & $ 10.32$  &  $  4.8\times 10^{-6}  $ \\   
 28 &     $105.0  $   & $194.65522$  & $27.17606$ & $ 1.9 $ &     $  3.1\times 10^{-14} $ & $194.65653 $&  $27.17589 $&   $3.9 $&    $ 7.8\times 10^{-17} $  & $ 4.2 \pm  3.8  $ &  $ 8.08   $ &  $ 9.72 $  & $ 9.52 $  &  $  4.8\times 10^{-6}  $ \\   
 31 &     $23.0  $   & $158.14528$  & $65.04093$ & $ 0.7 $ &     $  1.1\times 10^{-13} $ & $158.14563 $&  $65.04108 $&   $9.4 $&    $ 1.1\times 10^{-16} $  & $ 0.8 \pm  2.8  $ &   $ 7.19   $ &  $ 9.95 $  & $ 9.79 $  &  $  4.5\times 10^{-6}  $ \\   
 33 &     $42.8  $   & $180.59480$  & $29.86180$ & $ 2.0 $ &     $  5.1\times 10^{-13} $ & $180.59342 $&  $29.86145 $&   $0.4 $&    $ 1.0\times 10^{-16} $  & $ 4.5 \pm  1.8  $ &   $ 6.89   $ &  $ 10.00$  & $ 9.83 $  &  $  1.9\times 10^{-5}  $ \\   
 35 &     $13.0  $   & $181.94222$  & $43.12635$ & $ 0.7 $ &     $  2.8\times 10^{-13} $ & $181.94208 $&  $43.12643 $&   $1.4 $&    $ 2.7\times 10^{-17} $  & $ 0.5 \pm  1.0  $ &   $ 5.84   $ &  $ 9.59 $  & $ 9.39 $  &  $  9.7\times 10^{-5}  $ \\   
 36 &     $15.3  $   & $179.11786$  & $55.12564$ & $ 0.7 $ &     $  1.1\times 10^{-13} $ & $179.11730 $&  $55.12529 $&   $0.7 $&    $ 4.3\times 10^{-17} $  & $ 1.7 \pm  0.6  $ &   $ 6.47   $ &  $ 10.00$  & $ 9.72 $  &  $  2.8\times 10^{-5}  $ \\   
 38 &     $104.4  $   & $126.18034$  & $29.98986$ & $ 0.2 $ &     $  1.1\times 10^{-12} $ & $126.18044 $&  $29.98994 $&   $0.5 $&    $ 3.3\times 10^{-17} $  & $ 0.4 \pm  0.4  $ &  $ 6.42   $ &  $ 9.97 $  & $ 10.01$  &  $  5.4\times 10^{-3}  $ \\   
 39 &     $141.0  $   & $149.81165$  & $12.98786$ & $ 0.3 $ &     $  9.5\times 10^{-13} $ & $149.81155 $&  $12.98791 $&   $0.7 $&    $ 2.8\times 10^{-16} $  & $ 0.4 \pm  0.7  $ &  $ 7.78   $ &  $ 9.67 $  & $ 9.99 $  &  $  3.7\times 10^{-4}  $ \\   
 40 &     $151.5  $   & $220.15871$  & $53.50445$ & $ 0.3 $ &     $  1.6\times 10^{-12} $ & $220.15876 $&  $53.50438 $&   $1.7 $&    $ 8.1\times 10^{-16} $  & $ 0.3 \pm  0.6  $ &  $ 8.22   $ &  $ 10.00$  & $ --   $    &  $  2.7\times 10^{-4}  $ \\   
 49 &     $30.9  $   & $139.99161$  & $37.19108$ & $ 0.3 $ &     $  3.9\times 10^{-12} $ & $139.99157 $&  $37.19101 $&   $0.3 $&    $ 4.0\times 10^{-17} $  & $ 0.3 \pm  0.3  $ &   $ 5.56   $ &  $ 9.91 $  & $ 10.01$  &  $  1.2\times 10^{-2}  $ \\   
 50 &     $4.4  $   & $186.45360$  & $33.54686$ & $ 0.2 $ &     $  6.3\times 10^{-12} $ & $186.45383 $&  $33.54680 $&   $0.5 $&    $ 1.7\times 10^{-17} $  & $ 0.7 \pm  0.4  $ &    $ 4.07   $ &  $ 9.10 $  & $ 12.16$  &  $  1.1\times 10^{-2}  $ \\   
 51 &     $197.7  $   & $191.64686$  & $ 2.36912$ & $ 0.2 $ &     $  1.9\times 10^{-12} $ & $191.64712 $&  $ 2.36929 $&   $2.2 $&    $ 2.7\times 10^{-17} $  & $ 1.1 \pm  1.9  $ &  $ 6.39   $ &  $ 9.84 $  & $ --   $    &  $  2.4\times 10^{-1}  $

 \end{longtable}
 \normalsize
 \renewcommand{\thefootnote}{\arabic{footnote}}
\renewcommand{\arraystretch}{1.0}
\end{landscape}
\twocolumn
}

\begin{figure*}
\centering
        \subfigure{%
            \includegraphics[width=0.43\columnwidth]{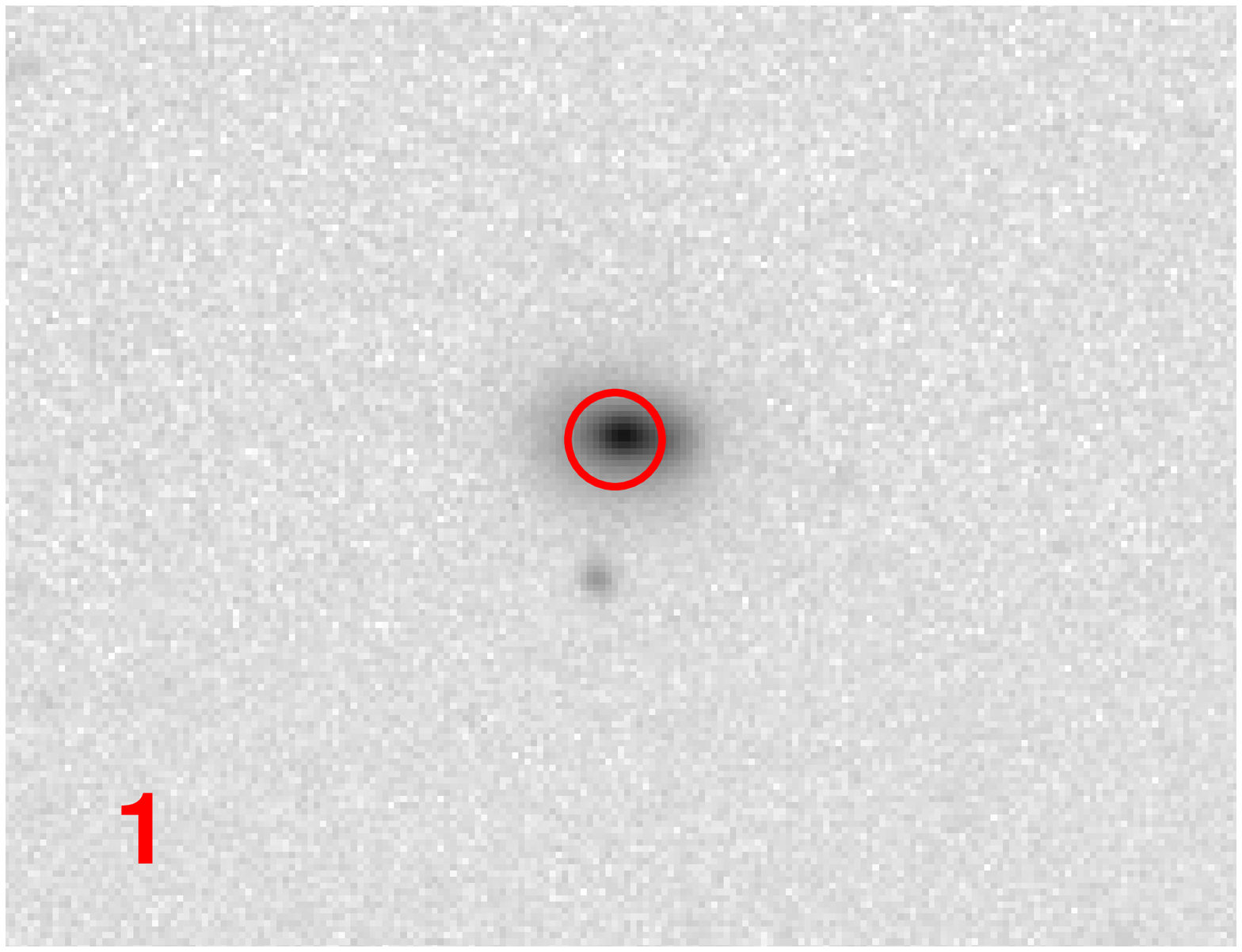}
        }
        \subfigure{%
            \includegraphics[width=0.43\columnwidth]{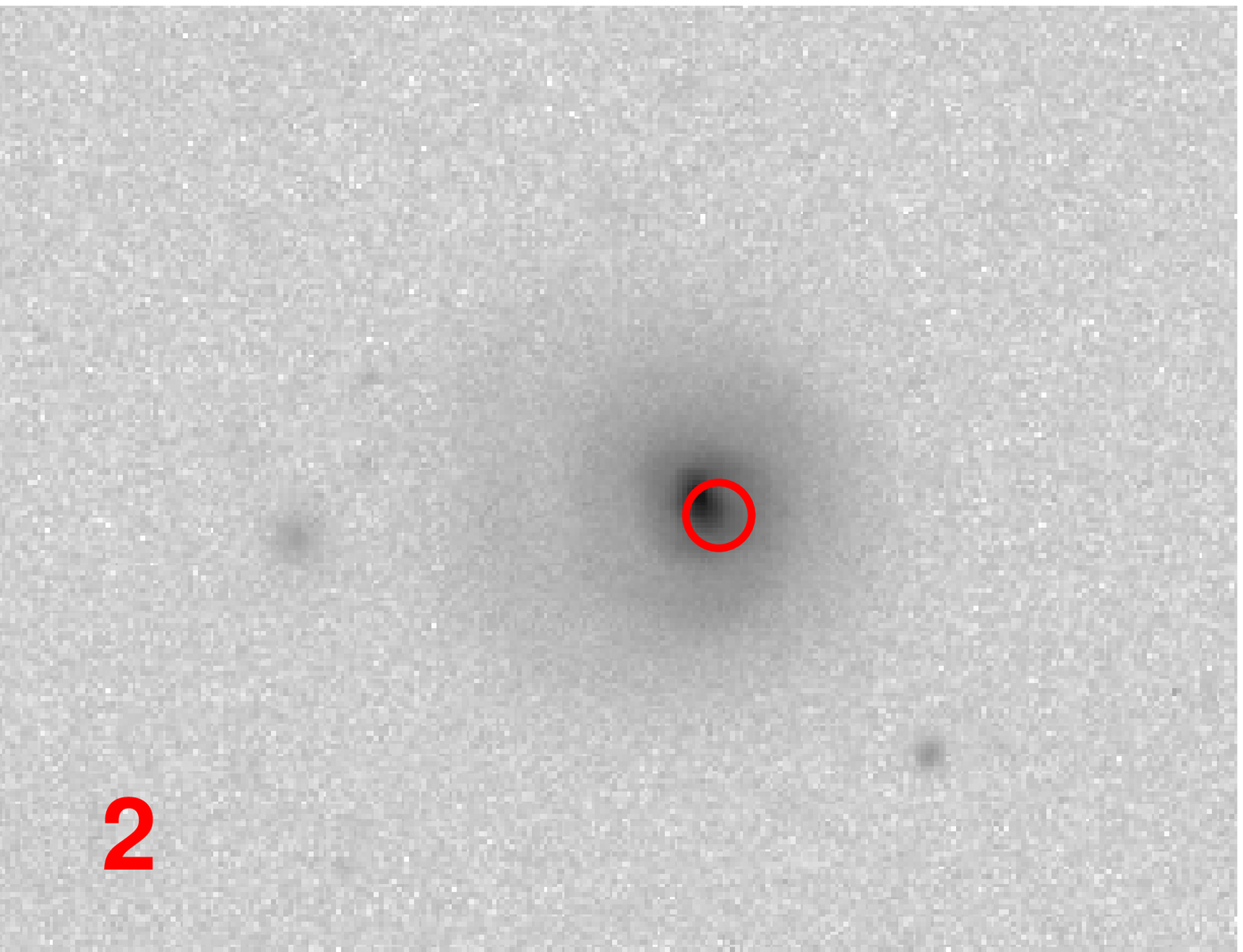}
        }
        \subfigure{%
            \includegraphics[width=0.43\columnwidth]{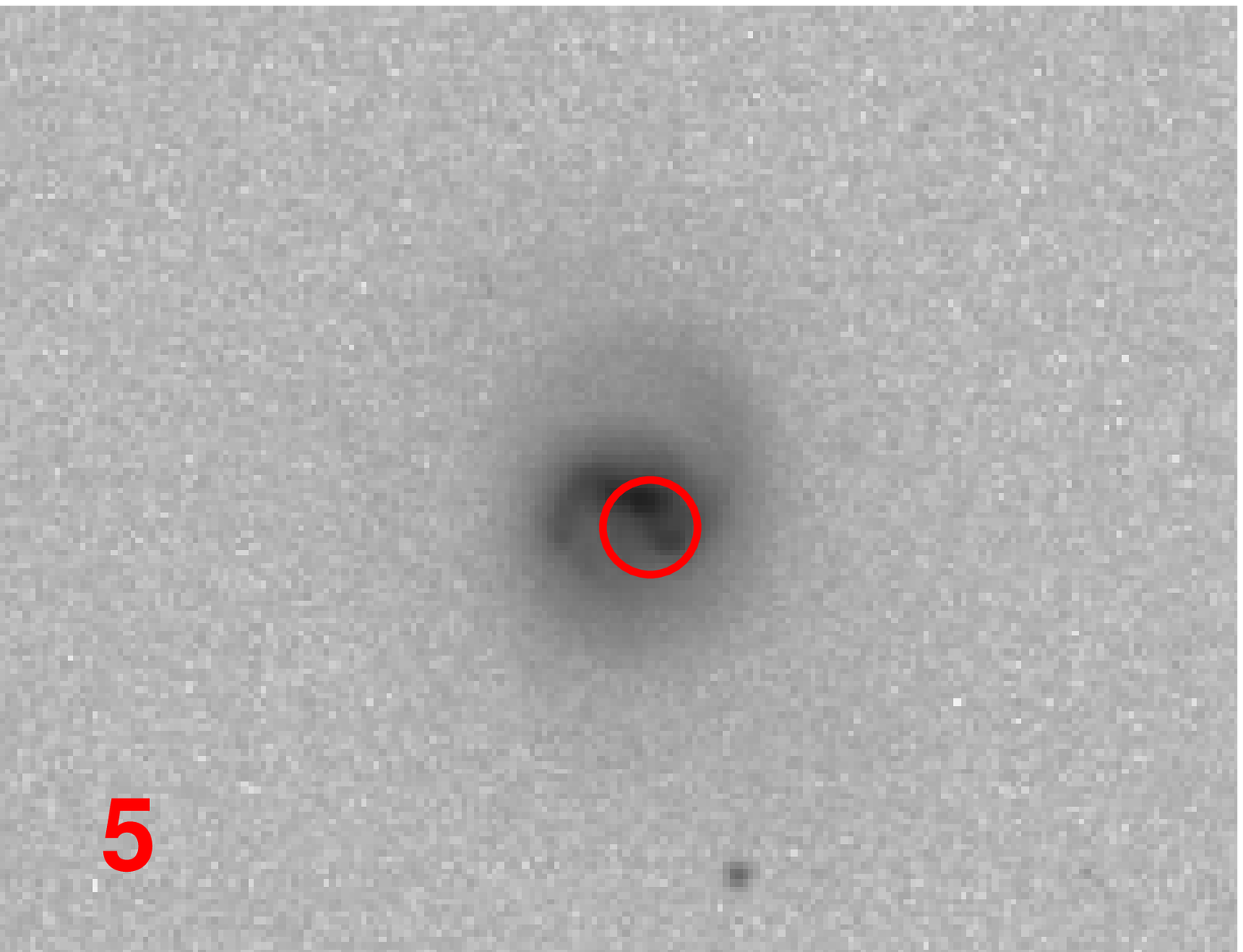}
        }\\
        \subfigure{%
            \includegraphics[width=0.43\columnwidth]{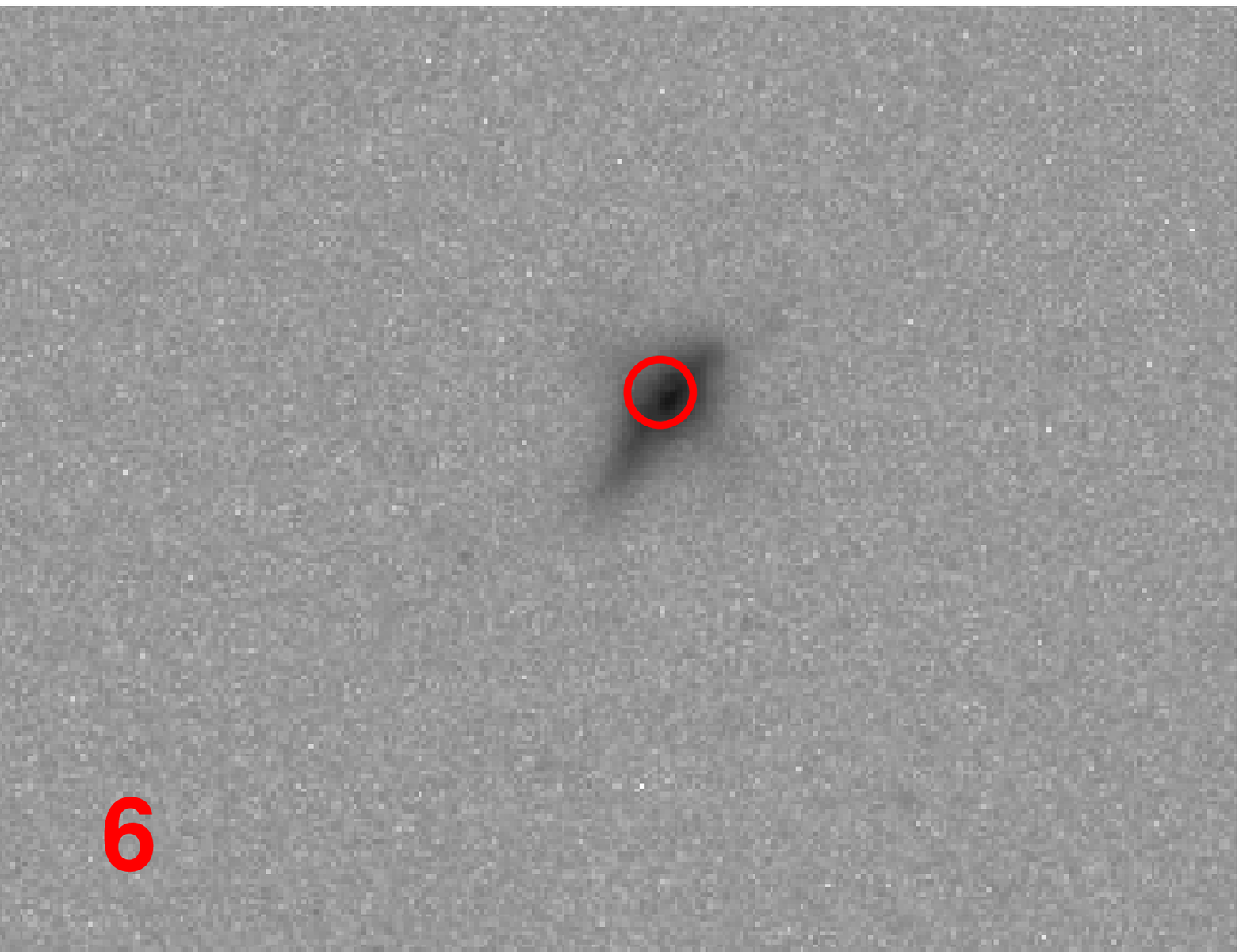}
        }
        \subfigure{%
           \includegraphics[width=0.43\columnwidth]{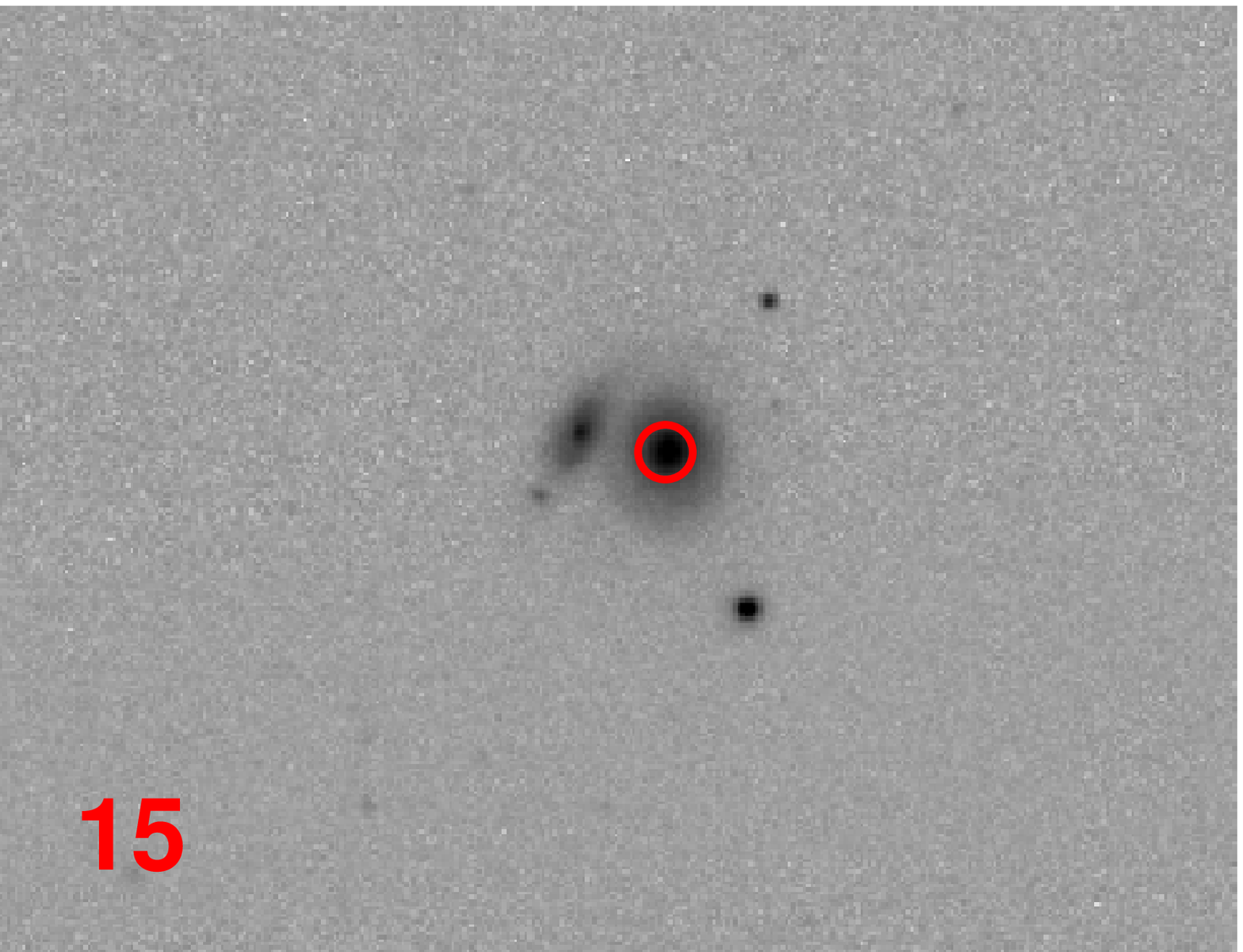}
        }
        \subfigure{%
            \includegraphics[width=0.43\columnwidth]{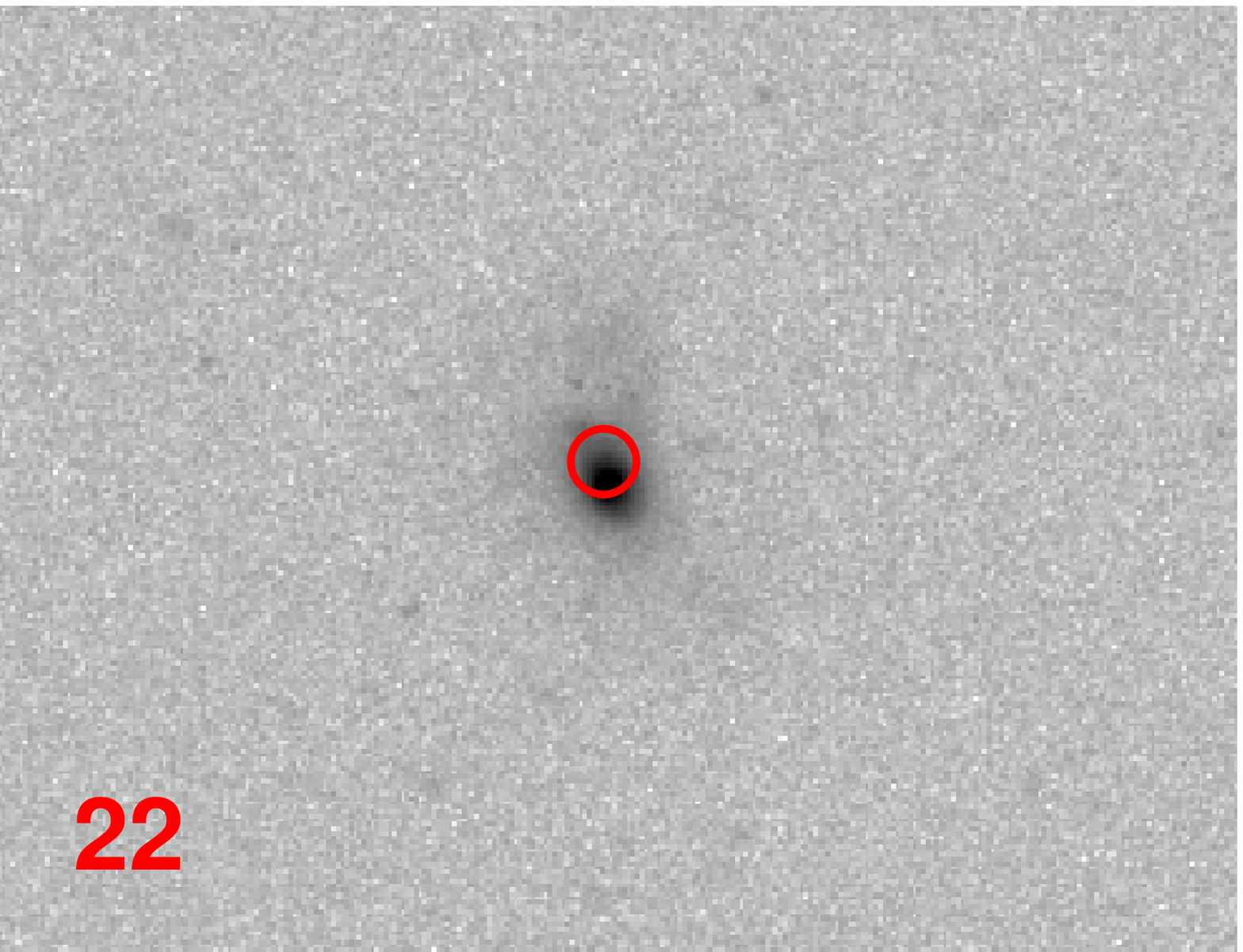}
        }\\
        \subfigure{%
            \includegraphics[width=0.43\columnwidth]{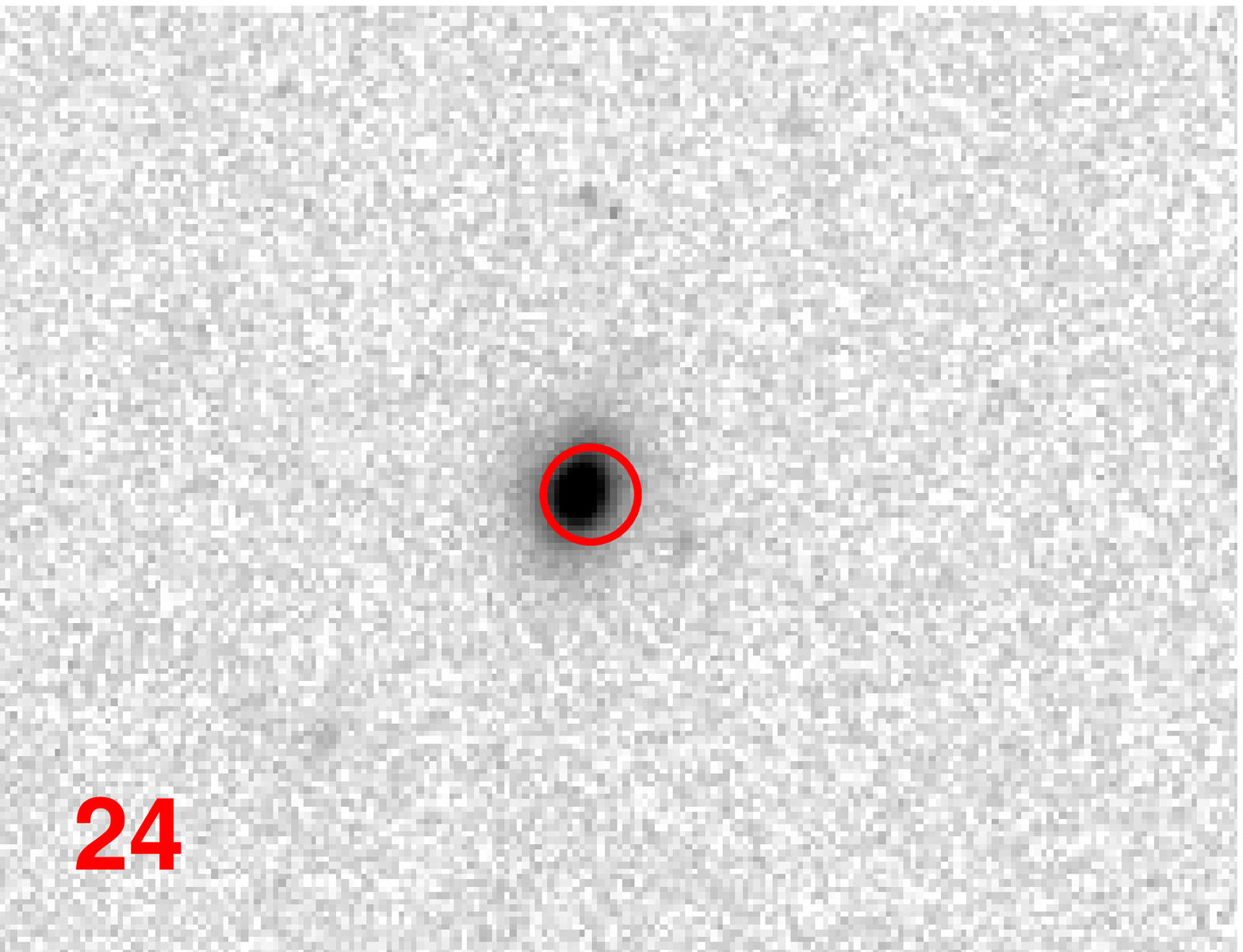}
        }
        \subfigure{%
            \includegraphics[width=0.43\columnwidth]{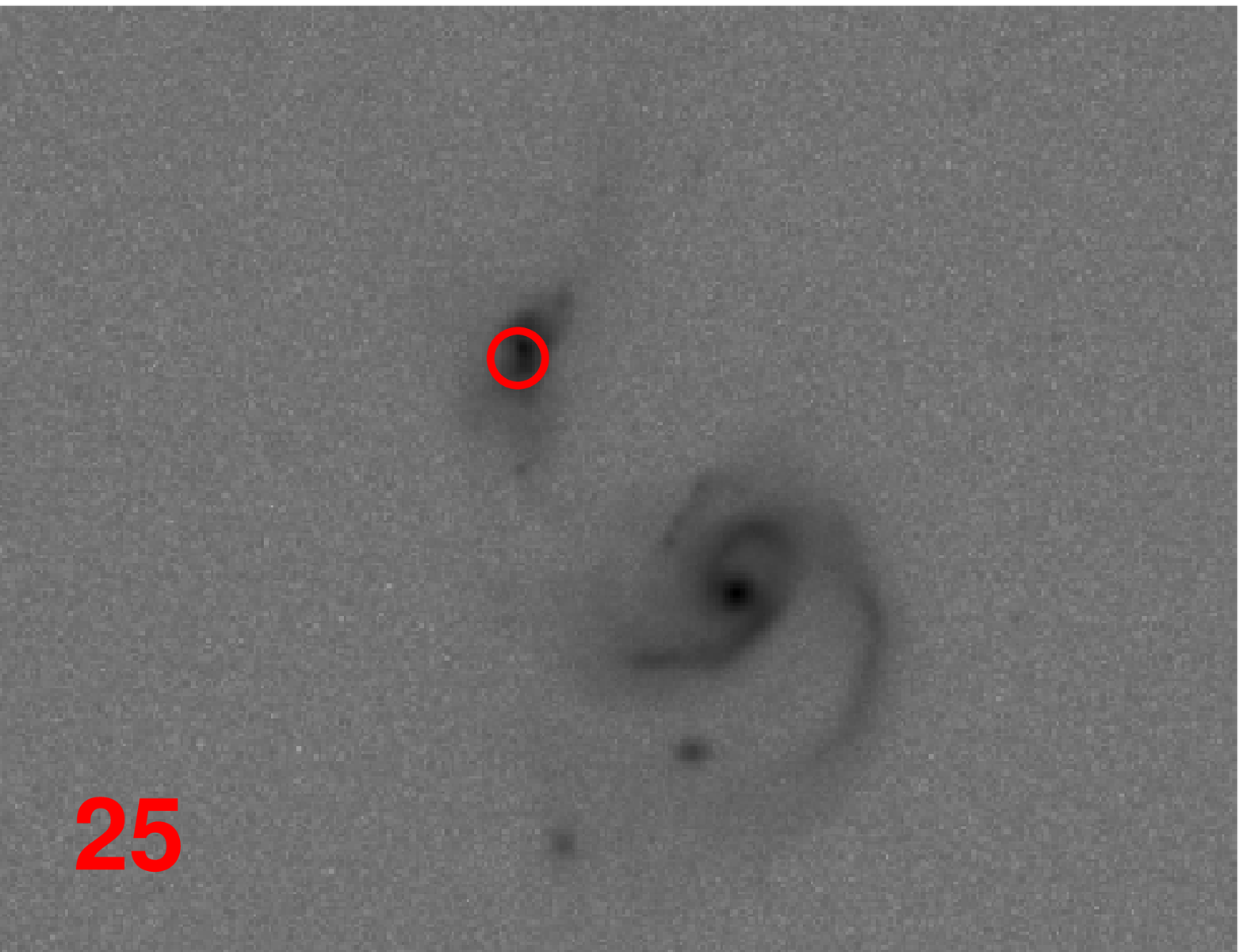}
        }
        \subfigure{%
            \includegraphics[width=0.43\columnwidth]{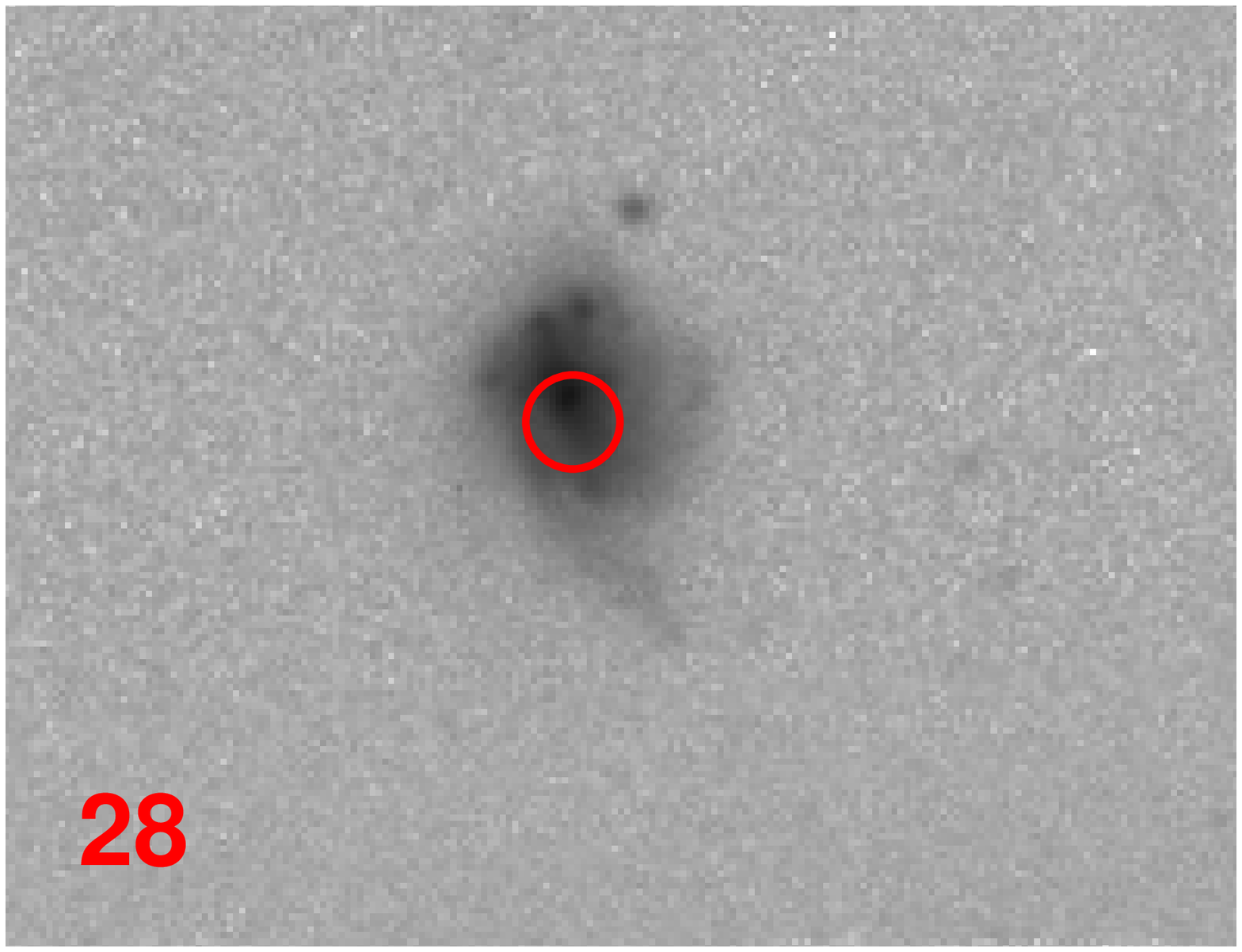}
        }\\
        \subfigure{%
             \includegraphics[width=0.43\columnwidth]{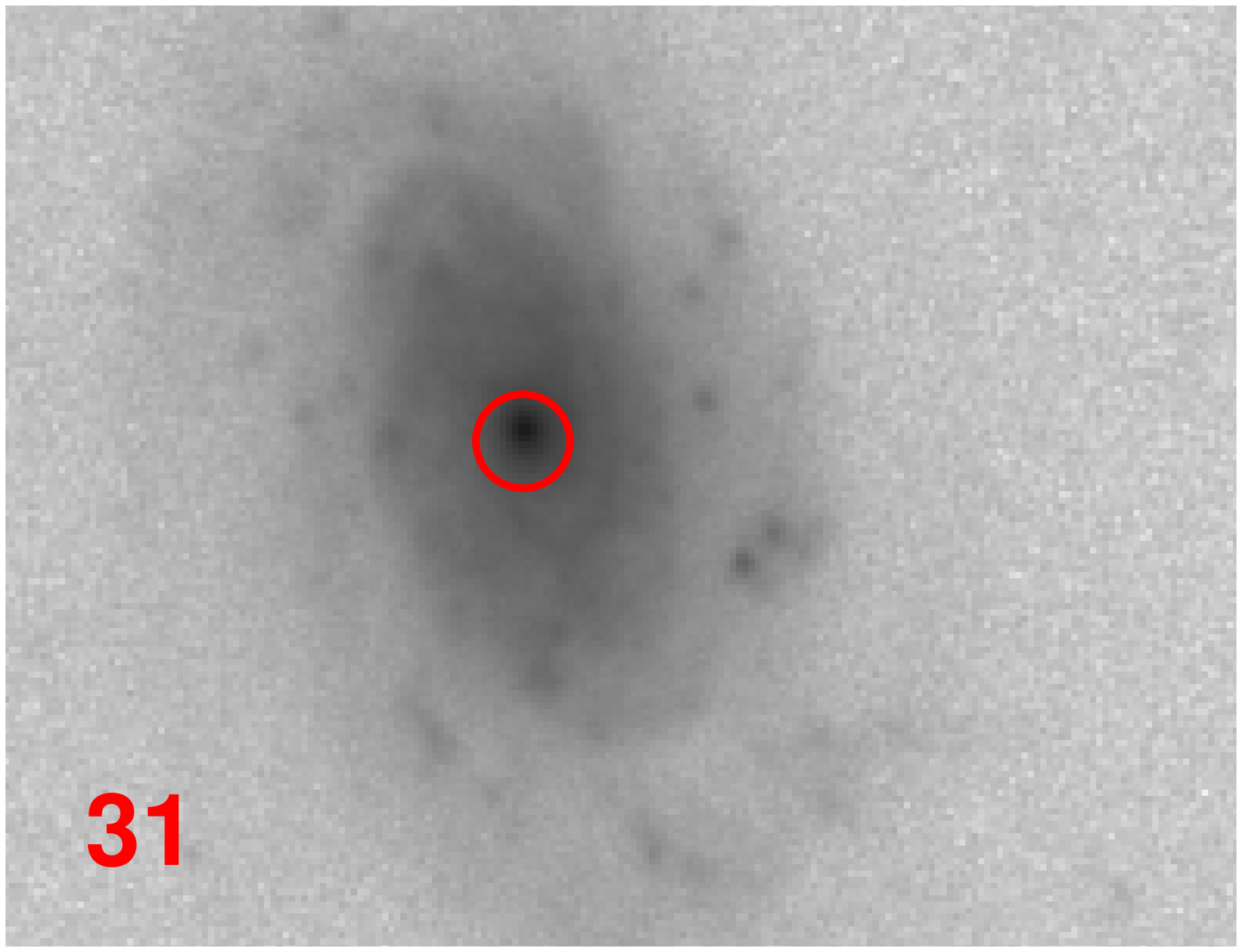}
        }
        \subfigure{%
                          \includegraphics[width=0.43\columnwidth]{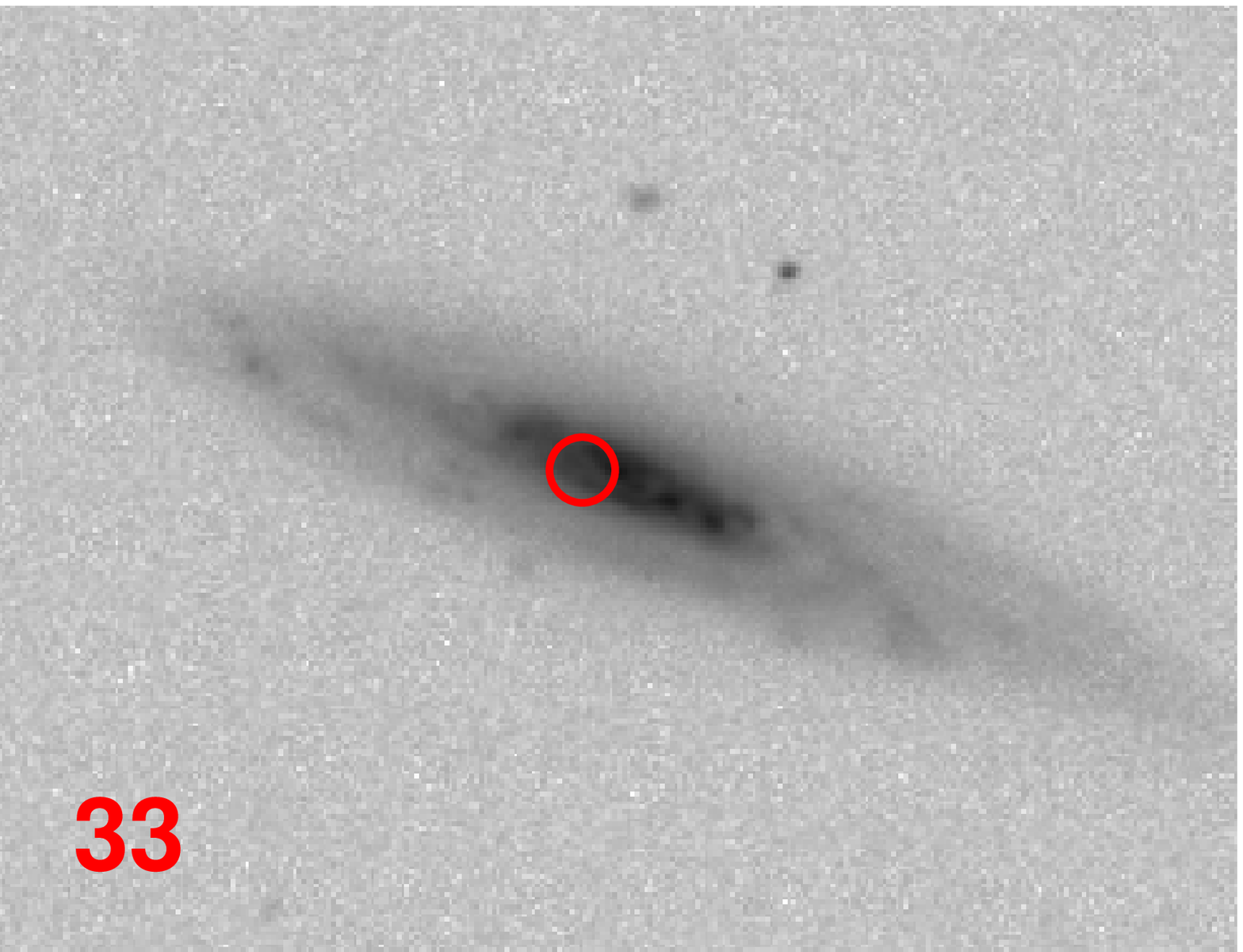}
        }
        \subfigure{%
            \includegraphics[width=0.43\columnwidth]{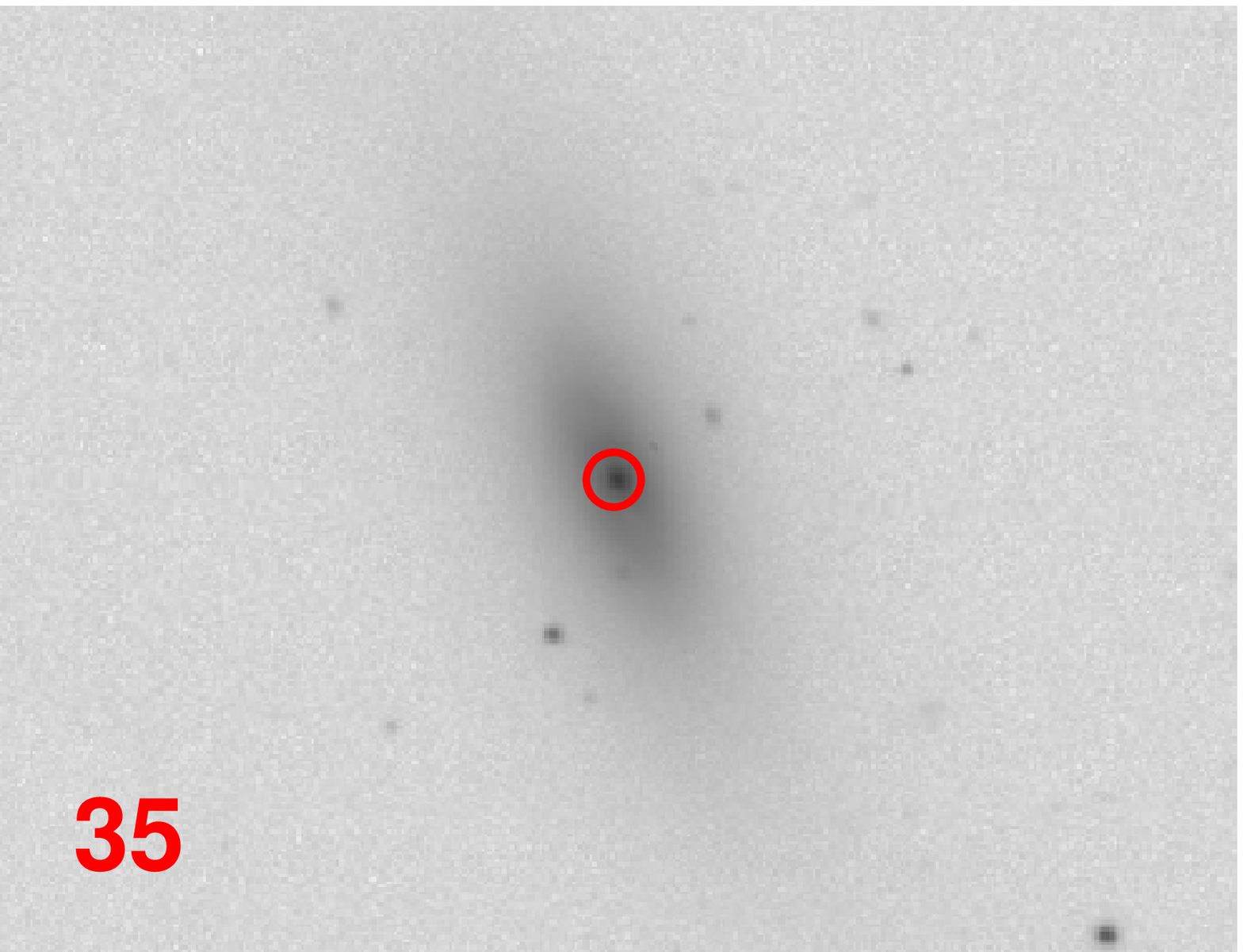}
        }\\
        \subfigure{%
            \includegraphics[width=0.43\columnwidth]{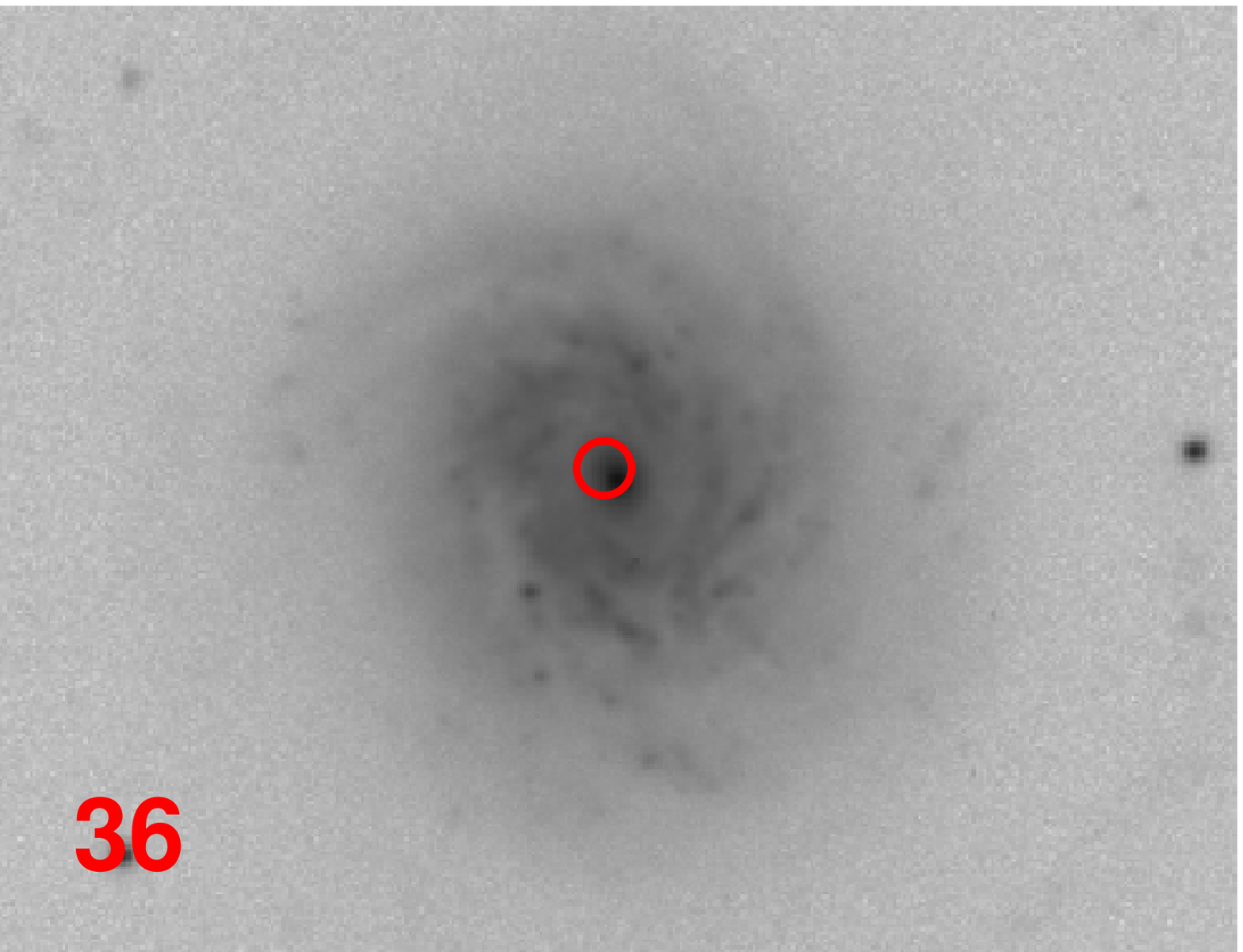}
        }
        \subfigure{%
            \includegraphics[width=0.43\columnwidth]{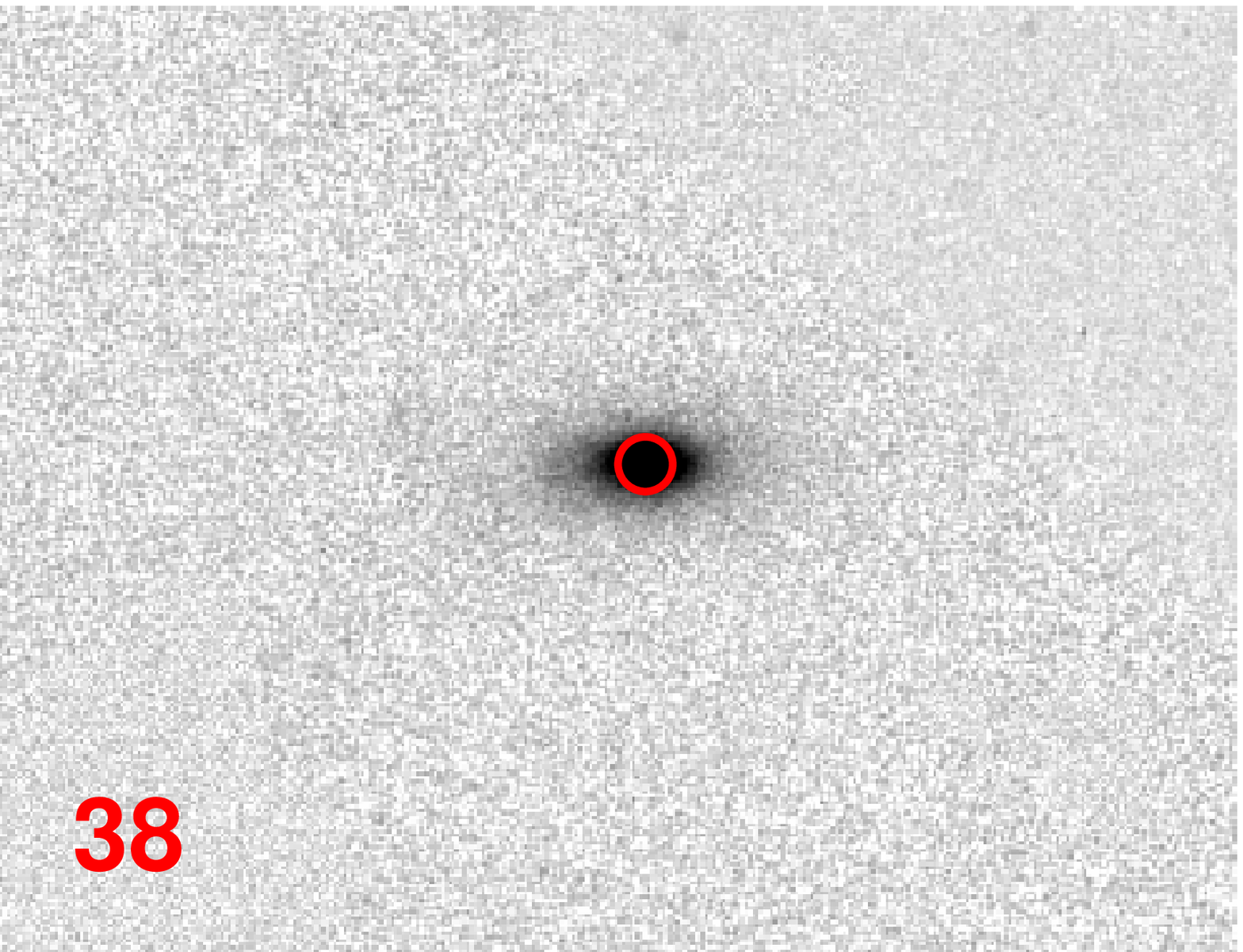}
        }
        \subfigure{%
           \includegraphics[width=0.43\columnwidth]{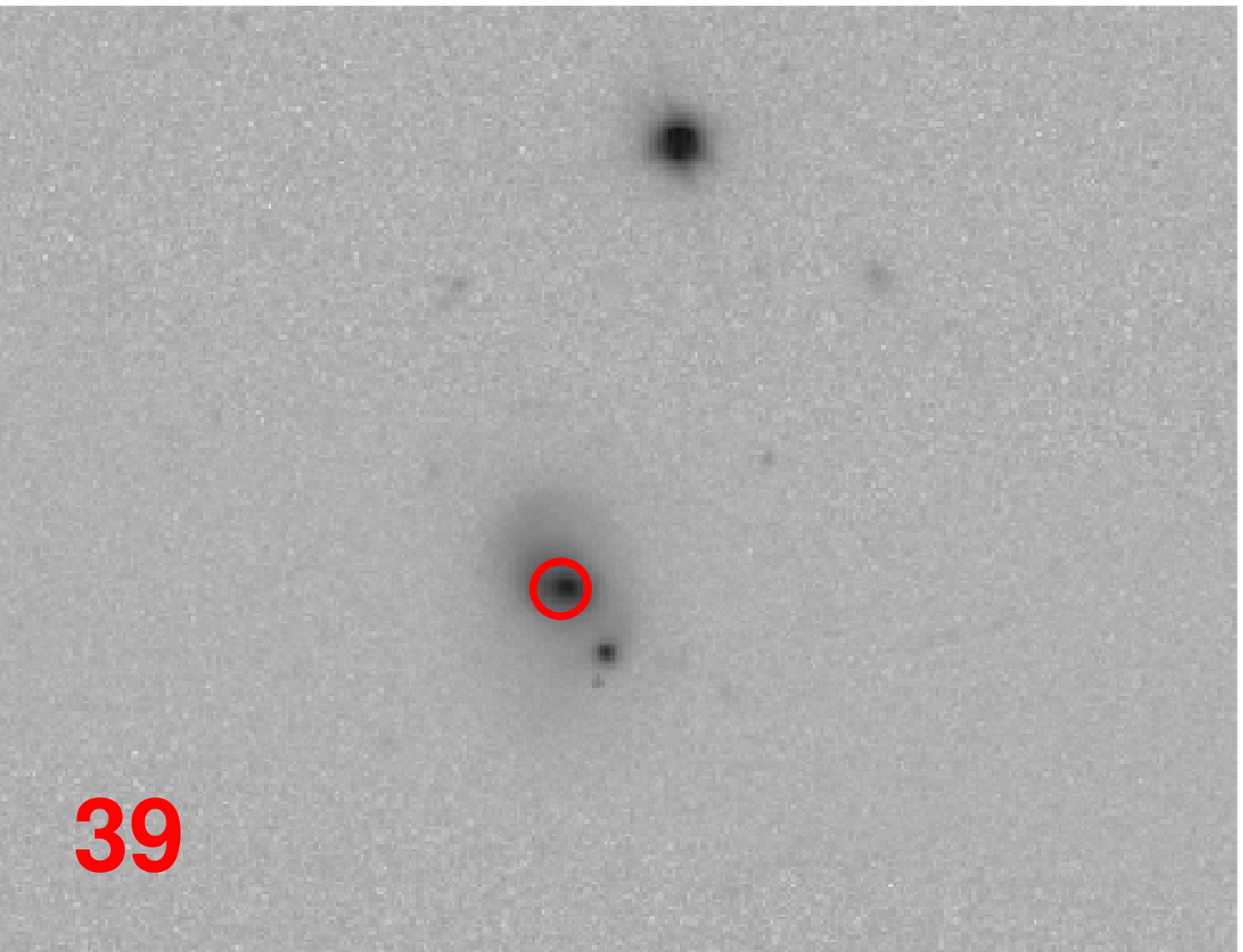}
        }\\
        \subfigure{%
            \includegraphics[width=0.43\columnwidth]{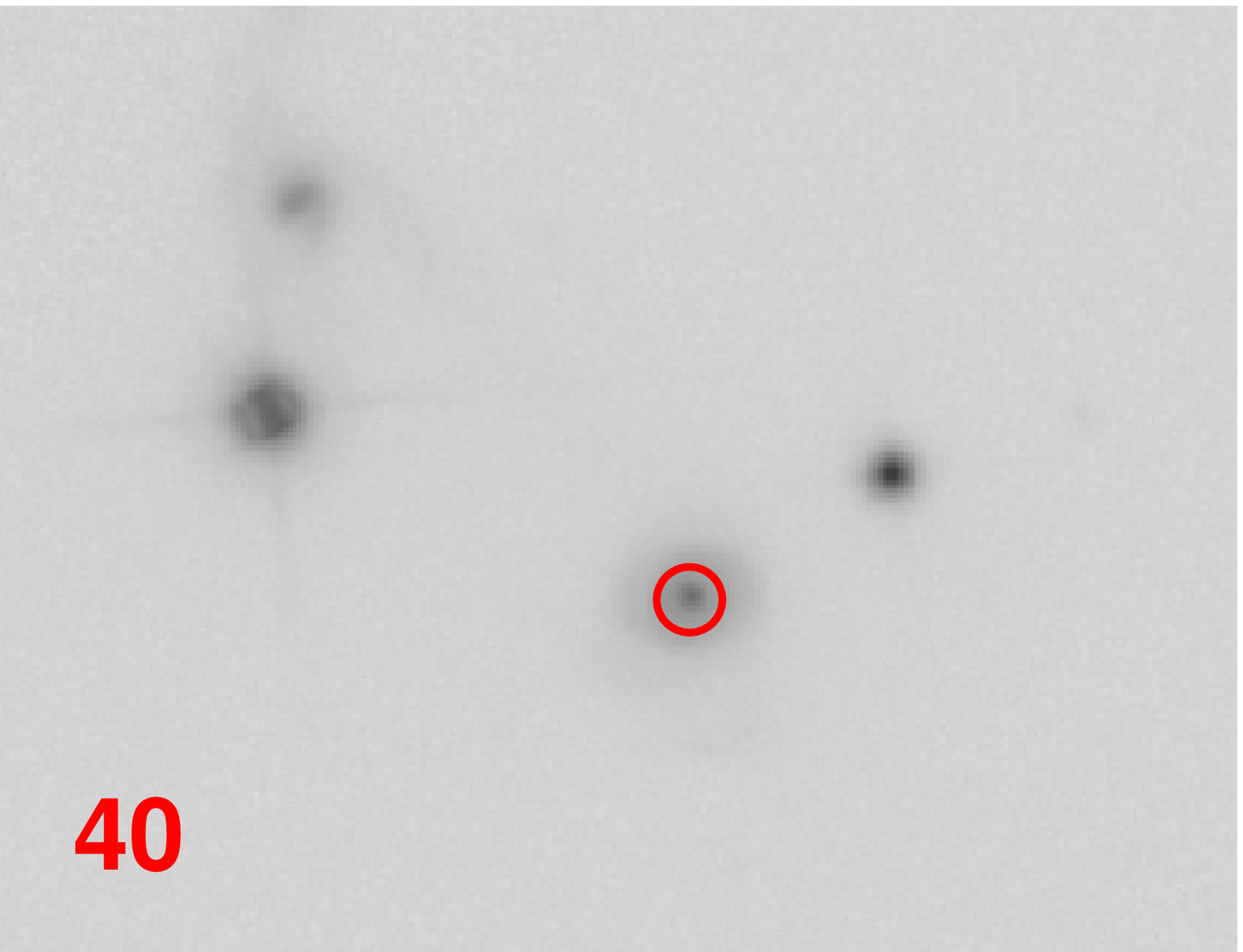}
        }
        \subfigure{%
            \includegraphics[width=0.43\columnwidth]{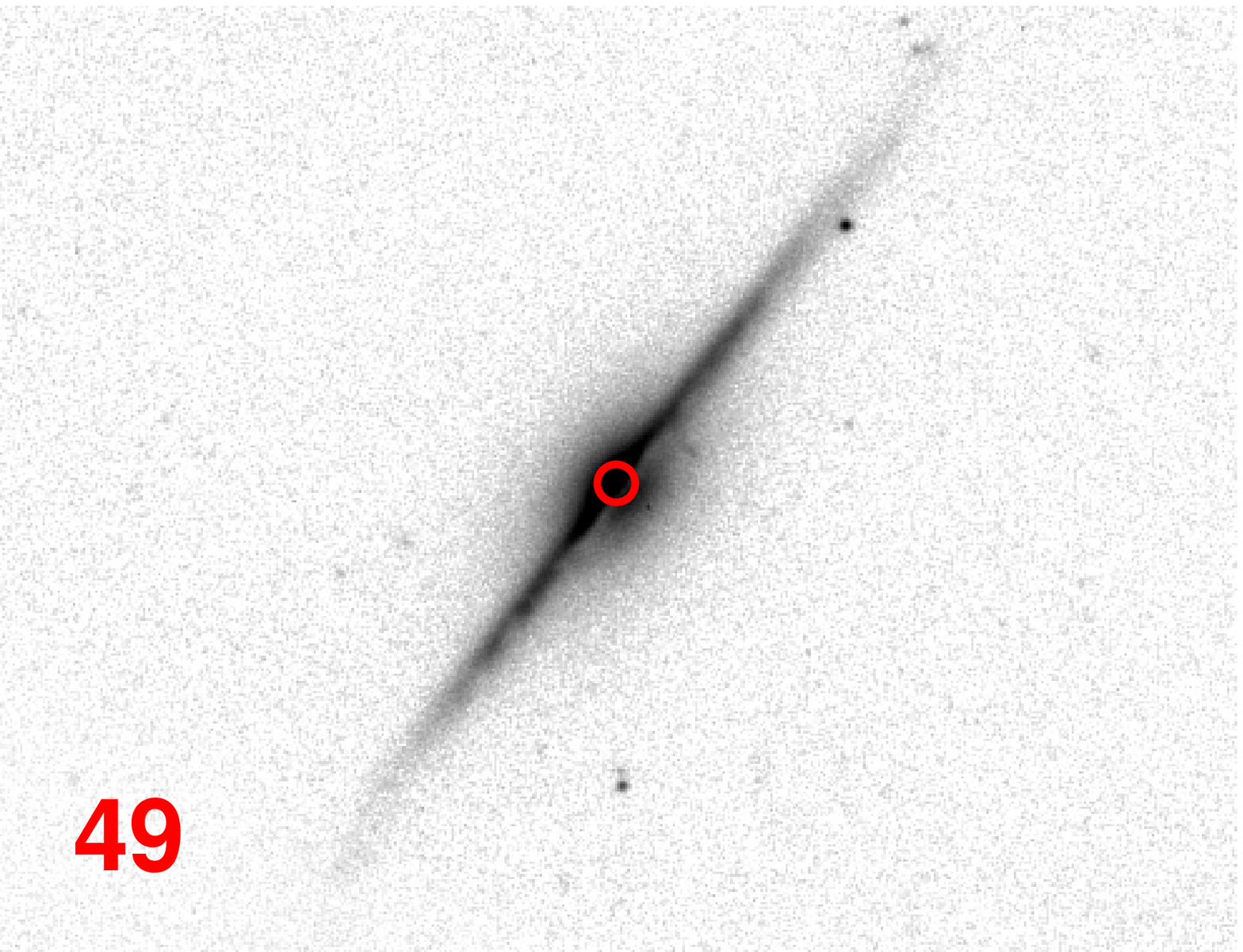}
        }
        \subfigure{%
            \includegraphics[width=0.43\columnwidth]{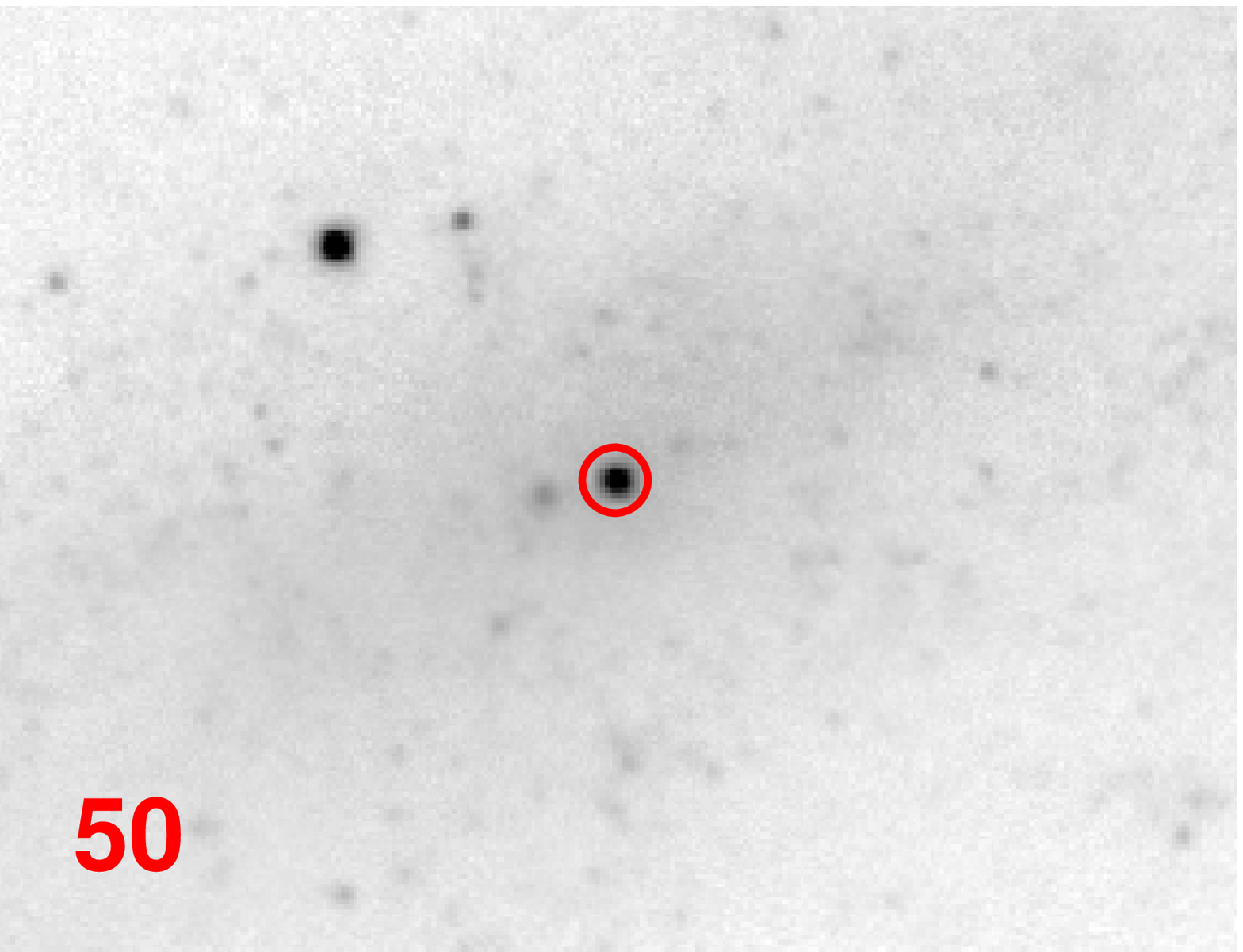}
        }\\
        \subfigure{%
           \includegraphics[width=0.43\columnwidth]{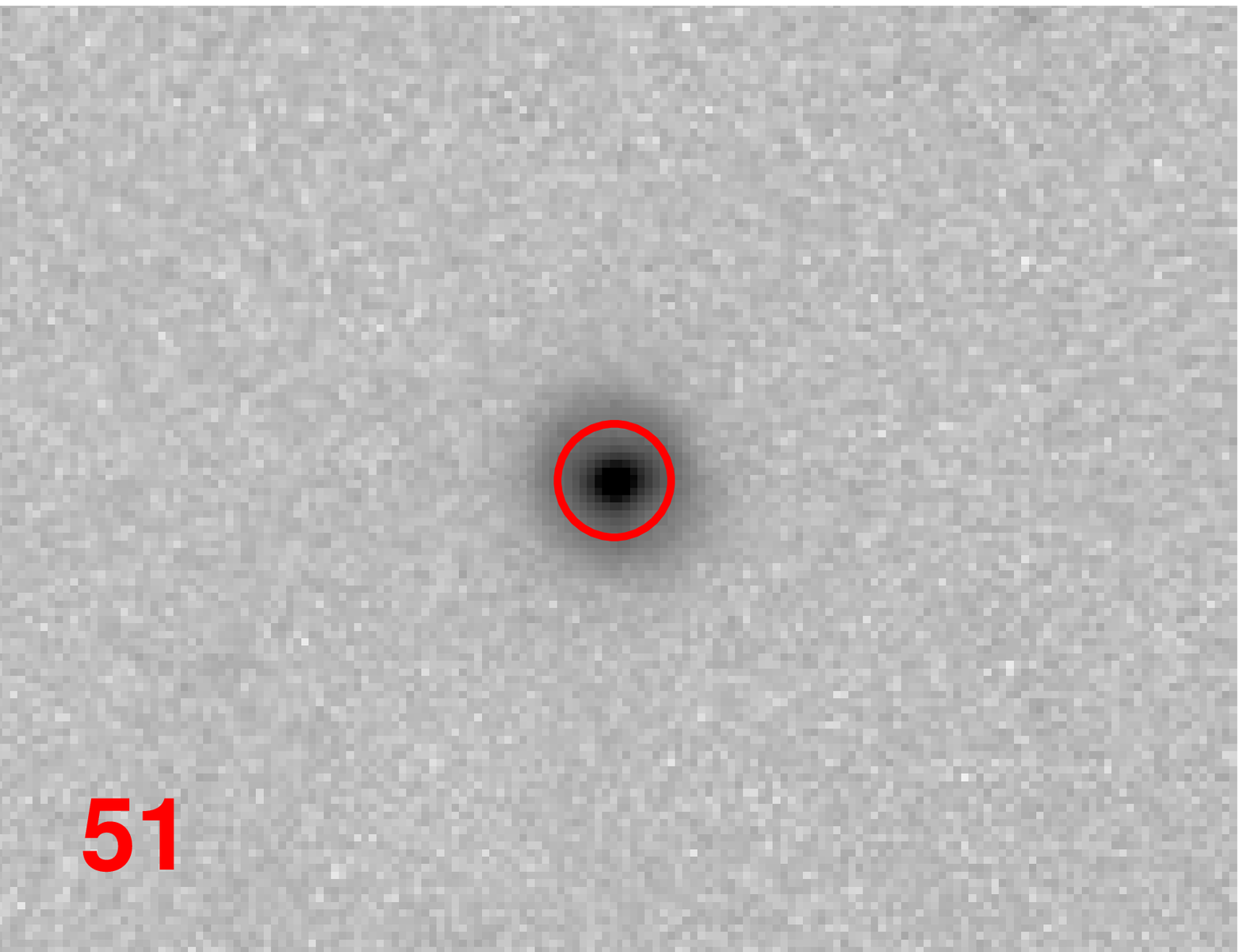}
        }
 \caption{SDSS images in the {\it r} band of the sub-sample of our GiX catalogue showing evidence of nuclear sources emitting in $X$-ray and 
radio. The red circle (having a radius of $3\arcsec$) represents the position of the central candidate black hole and is
 positioned on the associated $X$-ray source coordinates.}
   \label{figfinal}
\end{figure*}

% % xxxxxxxxxxxxxxxxxxxxxxxxxxxxxxxxxxxxxxxxxxxxxxx
\begin{figure*}
\centering
\hspace*{-1.25cm}
{\includegraphics[width=2.0\columnwidth, ]{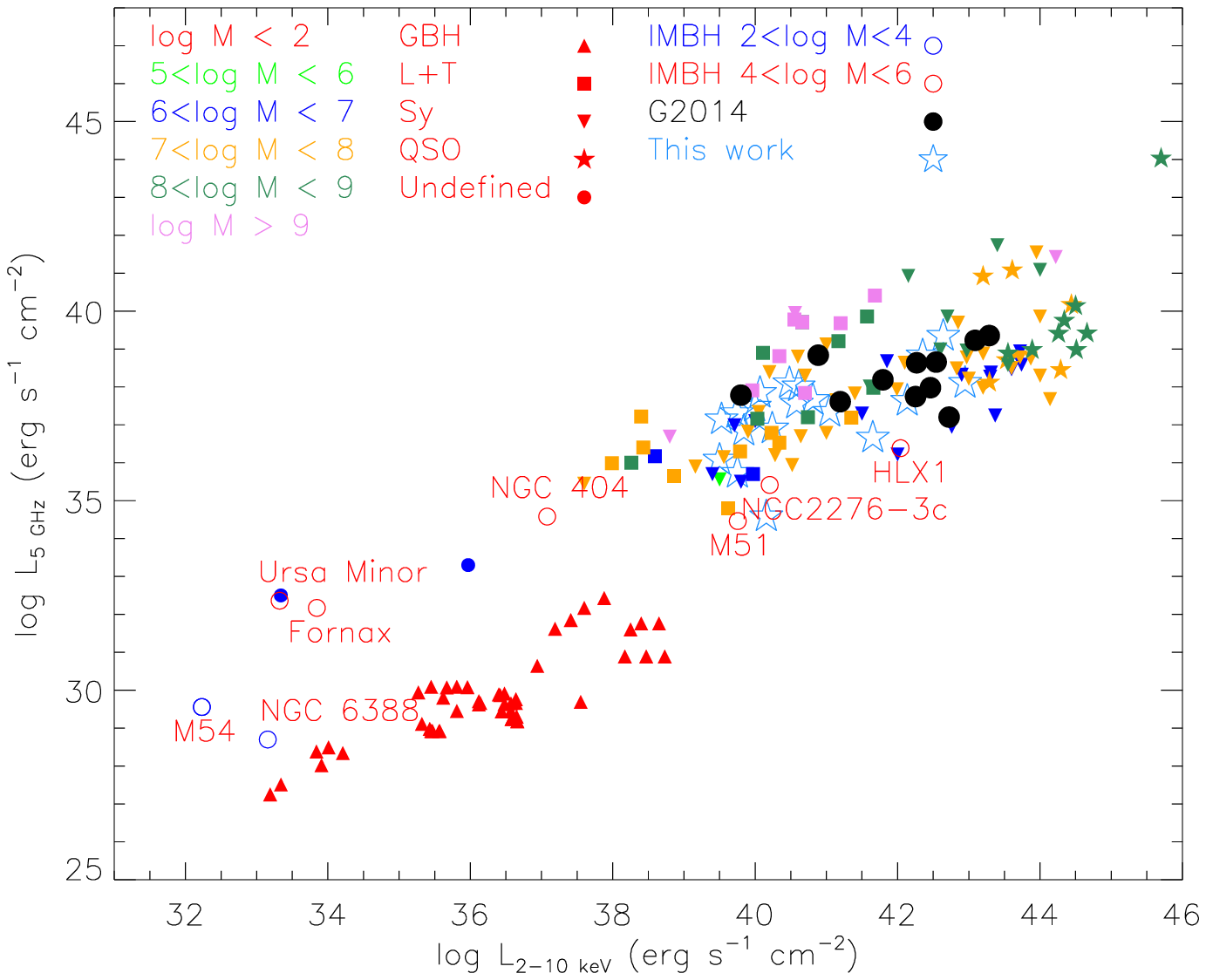}}
%{\includegraphics[width=1.2\columnwidth, ]{merloni.eps}}
\caption{The fundamental plane adapted from \citet{merloni2003} super-imposing also our results (dodger blue empty stars). 
We also consider Galactic BHs (GBH), Liners-Transition and quasi stellar objects (L-T and QSO, respectively), Seyfert nuclei (Sy), undefined sources, as well as IMBH candidates (open circles), 
already known in literature (see text for details). 
Filled black circles represent eleven low-mass AGN showing evidence of IMBHs with mass less than $10^{6.3}$ M$_{\odot}$ 
(\citealt{gultekin2014}).}
%\caption{The fundamental plane adapted from \citet{merloni2003} super-imposing also our results (dodger blue empty stars). We also consider the intermediate 
%mass black hole candidates (open circles) already known in literature 
%(see e.g. \citealt{nucita2008,wrobel2011,webb2012,nyland2012,nucita2013a, nucita2013b, gultekin2014,manni2015,mezcua2015,earnshaw2016a,earnshaw2016b}). These IMBHs are 
%possibly hosted in selected globular clusters and nearby dwarf galaxies and the associated mass is determined via dynamical methods. 
%Filled black circles represent eleven low mass AGN showing evidence of IMBHs with mass less than $10^{6.3}$ M$_{\odot}$ (\citealt{gultekin2014}).}
\label{figmerloni}
\end{figure*}
\section{Conclusions}
\label{results}
{
In this paper we presented {\it GiX}, a sample of $X$-ray detected low mass galaxies with mass $\ut < 10^10$ M$_{\odot}$. 
The catalogue has been obtained by cross-correlating the NASA-Sloan 
Atlas with the 3XMM catalogue and consists of 51 $X$-ray emitting low mass galaxies. {We also matched the catalogue with the FIRST, NVSS and RG2007}
databases and found 19 sources with a radio counterpart. Although dedicated follow up observations would be required in order to disentangle the nature of the sources, 
the targets, which are well consistent in position with the galactic center regions, are characterized by {  an $X$-ray} and radio luminosities that  
distribute according to the fundamental plane relation of \citet{merloni2003}. This fact allows one to use the previous relation (which has primarily a statistical nature) 
as a tool to estimate the mass of the 
central black hole candidates. Of course, the error associated to the estimated mass cannot be smaller than {that ($\pm 1.06$) derived by \citet{merloni2003} (to which we refer for more details) 
using the intrinsic scatter (i.e. the dispersion of the interesting variables) of the fundamental plane itself.}

{When performing this calculation, we found black hole candidates with mass in the range $\simeq 10^4-2\times 10^8$ M$_{\odot}$ (see the seventh column in Table \ref{tablemerloni}),
with eight candidates having mass below $10^{7}$ M$_{\odot}$. 

By comparing the observed $0.2-12$ keV luminosity with the 
Eddington limit, we found that all the sources appear to radiate with efficiency in the range $\simeq 0.2-10^{-6}$, with only 3 sources over 19 emitting with very high efficiency ($\epsilon_X \ut> 10^{-2}$) 
and 3 black hole candidates having intermediate values of efficiency in the range $10^{-2}-10^{-4}$.}

A similar analysis has been recently performed by \citet{lemons2015} by cross-correlating a sample of dwarfs with the Chandra Source Catalogue and finding 19 
galaxies with a number of point-like $X$-ray sources with hard spectra and $2-10$ keV luminosities between 
$10^{37}$ erg s$^{-1}$ and $10^{40}$ erg s$^{-1}$, i.e. in the typical range of power emitted by stellar-mass $X$-ray 
binaries and massive black holes accreting at relatively low Eddington rate. 

{By using the data reported in Table \ref{tablemerloni}, and having in mind the large uncertainties due to the intrinsic scatter of the fundamental plane, 
it is interesting to note that {7 targets} have $M_{gal}/M_{BH}$ in the range $[1,100]$, 5 targets have $M_{gal}/M_{BH}$ in the range $]100,1000]$, while 
7 entries have $M_{gal}/M_{BH}$ {  larger than or equal to 1000}. In the first case, the galaxies seem to be outliers in the black hole-host galaxy scaling relations. This is particularly true 
for source $\# 1$ (NGC 1303) for which the galaxy to black hole mass ratio is of the order 
of $\simeq 4$.} As a matter of fact, recently \citet{seth2014} (but see also \citealt{reines2014}) analyzed the adaptive optics kinematic data of 
M60-UCD1 (an ultra-compact dwarf galaxy with mass $\simeq (1.2\pm 0.4)\times 10^8$ M$_{\odot}$).
In order to explain the observed stellar velocities and the light distribution within the galaxy, these authors must assume the presence 
of a central massive black hole with mass of $2.1^{+1.2}_{-0.7}\times 10^7$ M$_{\odot}$, thus resulting in a galaxy to black hole mass ratio 
$M_{gal}/M_{BH}$ between $\simeq 2.3$ and $\simeq 11$ (with a central value of $\simeq 5.7$)\footnote{\citet{seth2014} explained their finding postulating that M60-UCD1 is the stripped nucleus of a galaxy due to the interaction
with the galaxy M60 which probably happened about 10 billions years ago.}. 
Although the discovery of such massive object is impressive,  
recent works (see, e.g., \citealt{reines2013}) already suggested the existence of massive black holes in dwarf galaxies when using virial techniques. Note however that these authors, 
using a sample of 151 dwarf galaxies that exhibit optical spectroscopic signatures of accreting black holes, found black holes candidates with masses in the range $10^5-10^6$ M$_{\odot}$ and 
median $2\times 10^5$ M$_{\odot}$, i.e. significantly smaller than those derived in this work using correlation between $X$-ray and radio data only. As a matter of fact, our sample is characterized by black hole candidates with mass in the range 
$10^{4.1}-10^{8.3}$ M$_{\odot}$ and median value $\simeq 3\times 10^7$ M$_{\odot}$. However, we remind that these values 
were estimated assuming a radio spectral index $\alpha_R=0$ from which the black hole mass slightly depends on (see Sect. 4).}

Of course, $X$-ray emitting sources in low-mass galaxies may have had a key role in the evolution of such systems, thus making urgent a 
theoretical background able to explain how such massive objects formed. In this respect, next coming optical and $X$-ray surveys, as those planned to be performed by 
Euclid and eRosita, will certainly help in solving this puzzling problem. Euclid, whose primary goal is to map the geometry of the dark universe 
and to constrain the dark energy content (see, e.g. \citealt{euclid2010} and \citealt{cimatti2012}) by performing a $\simeq 20000$ square degrees survey 
down to $\simeq 24.5$ AB magnitude, will allow also to detect (as a by-product) $\simeq 10^5$ dwarf galaxies up to $\simeq 100$ Mpc \citet{laureijs2011}. 
During the life-time of the Euclid mission, the eRosita (extended ROentgen Survey with an Imaging Telescope Array) will be fully operational on-board the 
Russian Spektrum-Roentgen-Gamma (SRG) mission (see \citealt{merloni2012}). 
This instrument will have a spatial resolution on axis comparable to that of the {\it XMM}-Newton satellite but with larger effective area.
After the completion of the four year all-sky survey, an average exposure of $2548$ s per FOV ($0.83$ deg$^2$) will allow one to detect rather faint objects with flux limits 
of $\sim 10^{-14}$ erg s$^{-1}$ cm$^{-2}$ and $\sim 10^{-13}$ erg s$^{-1}$ cm$^{-2}$ in the energy bands $0.5-2$ keV and $2-10$ keV, respectively. 
Hence, the combination of the two missions could result in the discovery of a remarkably high number of $X$-ray sources in low-mass galaxies up to moderate 
large redshifts. Further dedicated radio observations towards the selected targets will then permit to investigate 
the massive black hole properties in such galaxies. Indeed, as noted by \citet{mezcua2016}, determining the fraction of black holes in low-mass galaxies and the galaxy properties 
(as stellar density and star formation rate) at different redshifts could allow one to understand whether massive black holes form from a first generation of $\simeq 100$ M$_{\odot}$ stellar seeds 
or from the growing of {less massive} black holes (with $\simeq 10^4-10^6$ M$_{\odot}$) originated by the direct collapse of gas clouds in primordial halos (see, e.g. \citealt{pacucci2016}).

If massive black holes are really so common in low-mass galaxies and ultra-compact dwarfs, this would have implications for the demographics of such objects. Based on their results about 
M60-UCD1, \citet{seth2014} point to the fact that the amount of massive black hole in the Universe must be much larger than commonly thought. Of course, at least 
for sources $\# 1$ (NGC 1303)  which shows a galaxy to black hole mass ratio very extreme, further follow-up observations 
both in $X$-rays and radio bands, as well as in the optical, are necessary in order to confirm (or reject) our results.

\section*{Acknowledgements}
This research has made use of data obtained from the 3XMM {\it XMM}-Newton serendipitous source catalogue compiled 
by the 10 institutes of the {\it XMM}-Newton Survey Science Centre selected by ESA. We acknowledge the support by the INFN projects TAsP (Theoretical Astroparticle Physics Project) and EUCLID. 
We thank the anonymous Referee for the suggestions that greatly improved the paper.

\end{document}